\documentclass[11pt]{iopart}

\usepackage{verbatim}
\usepackage{graphicx}
\usepackage[usenames,dvipsnames,svgnames,table]{xcolor}
\usepackage[breaklinks=true,colorlinks=true,linkcolor=blue,urlcolor=blue,citecolor=blue]{hyperref}
\usepackage{rotating}
\usepackage{graphicx}
\usepackage{dcolumn}
\usepackage{bm}
\usepackage{epsfig}
\usepackage{hyperref}
\usepackage{ulem}
\usepackage{appendix}
\usepackage{iopams}
\usepackage{mwe}
\usepackage{subfig}
\expandafter\let\csname equation*\endcsname\relax
\expandafter\let\csname endequation*\endcsname\relax
\usepackage{amsmath}
\usepackage[usenames,dvipsnames,svgnames,table]{xcolor}

\begin{document}

\title {Interacting hadron resonance gas model in magnetic field and the fluctuations of conserved charges }
\author{\large Guruprasad Kadam$^1$, Somenath Pal$^2$ and Abhijit Bhattacharyya$^2$}
\ead{abhattacharyyacu@gmail.com}
\address{$^1$Department of Physics,
	Shivaji University, Kolhapur,
	Maharashtra-416004, India}
\address{$^2$Department of Physics,University of Calcutta, 92, A.P.C. Road, Kolkata-700009, India}

\begin{abstract}
In this paper we discuss the  interacting hadron resonance gas model in presence of a constant external magnetic field. The short range repulsive interaction between hadrons are accounted through van der Waals excluded volume correction to the ideal gas pressure. 
Here we take the sizes of hadrons as  $r_\pi$ (pion radius) $= 0$ fm, $r_K$ (kaon radius) $= 0.35$ fm, $r_m$ (all other meson radii) 
$= 0.3$ fm and $r_b$ (baryon radii) $= 0.5$ fm.
We analyse the effect of uniform background magnetic field on the thermodynamic properties of interacting hadron gas.  We especially discuss the effect of interactions on the behaviour of  magnetization of low temperature hadronic matter. The vacuum terms have been regularized using magnetic field independent regularization scheme.  We find that the magnetization of hadronic matter is positive which implies that the low temperature hadronic matter is paramagnetic. We further find that the repulsive interactions have very negligible effect on the overall magnetization of the hadronic matter and the paramagnetic property of the hadronic phase remains unchanged. We have also investigated the effects of short range repulsive interactions as well as the magnetic field  on the baryon and electric charge number susceptibilities of hadronic matter within the ambit of excluded volume hadron resonance gas model. 
\end{abstract}



\section{Introduction}

 The phase diagram of strongly interacting matter described by quantum chromodynamics (QCD) is one of the active topic of research  today. At low temperature ($T$) and low baryon chemical potential ($\mu_B$) strongly interacting matter consists of colorless hadrons, while at high temperature and/or high baryon density the fundamental degrees of freedom are colored quarks and gluons. At low baryon density and high temperature transition from hadronic phase to quark gluon plasma phase (QGP), as suggested by Lattice QCD (LQCD) simulation, 
 is actually an analytic cross over~\cite{Aoki:2006br,Borsanyi:2010bp,Borsanyi:2015waa,Petreczky:2012fsa,Ding:2015ona,Friman:2011zz}.
   At low temperature and high baryon density, hadron-QGP phase transition is of first order as shown in many QCD effective model calculations~\cite{Asakawa:1989bq,Barducci:1989eu,Barducci:1993bh,Berges:1998rc,Halasz:1998qr,Scavenius:2000qd,Antoniou:2002xq,Hatta:2002sj,ab1}. The first principle LQCD calculations are not available at high baryon density since the Euclidean formulation of the theory suffers from the sign problem~\cite{Hands:2001jn,Aoki:2005dt,Alford:2002ng}. If hadron-QGP phase transition is indeed first order, at low temperature and high baryon density, then we expect that the first order phase transition line should end at a critical point (CEP) as one moves towards the high temperature and low density regime where the analytic crossover phase transition starts. Locating this QCD critical point is one of the hot topic of research in the experimental and theoretical high energy physics (see Ref.~\cite{Bzdak:2019pkr} for a recent review on the status of QCD critical point).  
 
Relativistic heavy-ion collision (HIC) experiments provide a unique opportunity to empirically study the QCD phase diagram over wide range of $T$ and $\mu_B$ values. The evolution of the matter created in HIC experiments can be simulated using equations of relativistic hydrodynamics~\cite{Teaney:2000cw,Romatschke:2009im,Kolb:2003dz,Jacobs:2004qv}. In these hydrodynamic simulations the equation of state (EoS), as a function of control parameters $T$ and $\mu_B$, is the necessary input. In the off-central HICs a huge magnetic field, $B\sim m_{\pi}^2$ ($\sim 10^{18}$G) are created due to relativistic motion of charged particles~\cite{Skokov:2009qp,Bzdak:2011yy,Deng:2012pc}. This additional parameter $B$ can affect the equation of state and hence it can have significant impact on the overall structure of the phase diagram. For instance, magnetic field may induce interesting phenomena on QCD matter, $viz.$ chiral magnetic effect~\cite{Fukushima:2008xe}, magnetic catalysis~\cite{Shovkovy:2012zn}, inverse magnetic catalysis~\cite{Bali:2011qj,Preis:2012fh} etc. Further, strong magnetic fields are also expected to be present in dense neutron
stars~\cite{Duncan:1992hi,dey} and they may be present during the electroweak transition in the early universe~\cite{Vachaspati:1991nm,Bhatt:2015ewa}. The effect of magnetic field on transport phenomena are also very important in the context of HIC as well as neutron stars~\cite{Kadam:2014xka}. Thus it is of utmost importance to study the effect of magnetic field on the hot and dense strongly interacting matter.

Since the first principle lattice QCD calculations have limited applicability at finite baryon density one has to resort to effective models, $viz.$ 
Nambu-Jona-Lasinio model~\cite{Klevansky:1992qe,Hatsuda:1994pi}, quark-meson-coupling model~\cite{Schaefer:2007pw}, etc. The simplest effective model describing the hadronic phase of the strongly interacting matter is the hadron resonance gas (HRG) model. This model is based on the so called Dashen-Ma-Bernstein (DMB) theorem~\cite{Dashen:1969ep}.  It can be shown that if the dynamics of thermodynamic system of hadrons is dominated by narrow-resonance formation then the resulting system essentially behaves like a noninteracting system of hadrons and resonances~\cite{Dashen:1974yy,Welke:1990za,Venugopalan:1992hy}.  However, it fails to account for the short range repulsive interactions between hadrons. In fact, it has been shown that the repulsive interactions modelled via excluded volume corrections to ideal HRG partition function can have significant effect on thermodynamic observables, especially higher order fluctuations~\cite{zeeb,bugaev,vovchenko1,Albright:2015uua,Garg:2013ata,ab2} as well as in the context of statistical hadronization~\cite{BraunMunzinger:1999qy}.  HRG with repulsive interactions has also been used to study the transport coefficients of hadronic matter~\cite{Kadam:2015xsa,Kadam:2017iaz,Kadam:2019peo}.  Thus, it is necessary to include the repulsive interactions in the ideal HRG model if this model is to be used to understand the dynamics of hadronic matter in the context of HICs.

The study of correlations and fluctuations of conserved charges, $viz.$ baryon number, strangeness and electric charge, has recently gained a lot of attention.  These quantities are 
considered to be very sensitive probes of  phase transition~\cite{Deb,Lahiri,Datta,Majumder,Raha,Sur,Bhatta,repara,Sur1} in strongly interacting matter. Moreover, near the critical 
end point (CEP) the fluctuations are supposed to be large. Susceptibilities are related to fluctuations via the fluctuation-dissipation theorem. A measure of the intrinsic statistical fluctuations in a system close to thermal equilibrium is provided by the corresponding susceptibilities. 
The first principle LQCD simulations have been performed to compute susceptibilities at zero chemical potential. The susceptibilities are found to rise rapidly around crossover region of the phase diagram. Recently the susceptibilities have been estimated within the ambit of hadron resonance gas model (HRG)~\cite{Garg:2013ata} and its extended version, namely excluded volume hadron resonance gas model (EVHRG)~\cite{ab2}. While the Second order fluctuations and correlations estimated within ideal HRG model seem to agree reasonably with the lattice data, higher order fluctuations show deviations close to the transition temperature, $T_c$. It has been  argued that the breakdown of ideal HRG model near $T_c$ is the reflection of the fact that the hadrons melt quickly above $T_c$. However, in the recent study the higher order susceptibilities are found to be in very good agreement with lattice QCD data if the van der Waals interactions (attractive as well as repulsive) between baryons are included while keeping meson gas ideal~\cite{Vovchenko:2016rkn}. This study has concluded that the van der Waals interactions play very important role in describing the hadronic phase of QCD even near the QCD phase transition region.

In this work our purpose is to analyse  the effect of magnetic field on the EoS as well as on the conserved charge fluctuations in hot and dense hadronic matter using both HRG and EVHRG models. The presence of magnetic field not only affect the EoS~\cite{Endrodi:2013cs} but also the fluctuations and correlations~\cite{Bhattacharyya:2015pra}.  In the present study we have investigated the effect of magnetic field in the presence  of repulsive interaction on the EoS as well as on the susceptibilities of hadronic matter for the first time.  
 
We organize the paper as follows. In Sec.~\ref{secII} we recapitulate the thermodynamics of  hadron 
resonance gas model in magnetic field. In Sec.~\ref{secIII} we derive the renormalized vacuum pressure using magnetic field independent regularization (MFIR) scheme. In Sec.~\ref{secIV} we briefly discuss the extension of ideal HRG model to include repulsive interactions. In Sec.~\ref{secV} we discuss the results and finally in Sec.~\ref{secVI} we summarize and conclude.

\section{Hadron resonance gas model in magnetic field}
\label{secII}
The free energy density in the presence of constant external magnetic field $B$ is written in terms of partition function as 
\begin{equation}
F=-T\:\text{ln}\mathcal{Z}=F_{\text{vac}}+F_{\text{th}}
\end{equation}
where $F_{\text{vac}}$ and $F_{\text{th}}$ are vacuum and thermal parts respectively. In the ideal HRG model the free energy  at low temperature and in the dilute gas  approximation is given by the partition function of a gas of a non-interacting hadrons and resonances with very narrow spectral width~\cite{Dashen:1974yy,Welke:1990za,Venugopalan:1992hy}. This result is based on the Dashen-Ma-Bernstein theorem~\cite{Dashen:1969ep}. Thus, the free energy of non-interacting HRG model in presence of constant magnetic field is written as 

\begin{equation}
F_{c}=\mp\sum_{i}\sum_{S_z}\sum_{n=0}^{\infty}\frac{eB}{(2\pi)^2}\int dp_{z}\bigg(E_{i,c}(p_{z},n,S_{z})+T\:\text{ln}(1\pm e^{-(E_{i,c}-\mu_i)/T})\bigg); e_i\neq 0
\label{idFen}
\end{equation}

\begin{equation}
F_{n}=\mp\sum_{i}\sum_{S_z}\int\frac{d^3p}{(2\pi)^3}\:\bigg(E_{i,n}+T\:\text{log}\bigg[1\pm \text{exp}\bigg(-\frac{(E_{i,n}-\mu_i)}{T}\bigg)\bigg]\bigg); e_i=0
\end{equation}
\noindent
Here $e_i$ is the electric charge of $i^{\text{th}}$ hadronic species,
 $E_{i,c/n}$ is the single-particle
energy for charged/neutral particle, $m_i$ is the mass, $T$ is the temperature and $\mu_i=B_i\mu_B+\mathcal{S}_i\mu_\mathcal{S}+Q_i\mu_Q$ is the
chemical potential. In the expression of previous line, $B_i$, $\mathcal{S}_i$, $Q_i$, are
respectively, the baryon number, strangeness and electric charge of the particle,
$\mu_i^,$s are corresponding chemical potentials. The upper and lower
signs correspond to fermions and bosons respectively.
We have incorporated  the hadrons listed in the table~1.

For a constant magnetic field $B$, the single particle
energy levels for neutral and charged particles are respectively
given by~\cite{Endrodi:2013cs}

\begin{equation}
 E_{i,n}=\sqrt{p^2+m_i^2}
 \label{disp}
\end{equation}

\begin{equation}
 E_{i,c}(p_z,n,S_z)=\sqrt{p_{z}^2+m_i^2+2e_{i}B(n+1/2-S_z)}
 \label{dispmagn}
\end{equation}

\noindent
where, $n$ is any positive integer
corresponding to allowed Landau levels, $S_z$ is the component of spin
$S$ in the direction of magnetic field.
For a given $S$, there are $2S+1$ possible values of $S_z$. The
gyromagnetic ratios are taken as $g_i=2|e_i/e|$ ($e$ being elementary unit of electric charge) for all charged hadrons. It is to be noted that these values of gyromagnetic ratios are based on universal tree-level argument~\cite{Ferrara:1992yc}. Main reason for using this value is that the experimental value of $g$ is known only for few hadrons and the uncertainties in their values are also very large. The pressure of ideal hadron resonance gas  is $P^{\text{id}}=-F$ and it can be written as

\begin{equation}
P_{c}^{\text{id}}=\pm\sum_{i}\sum_{S_z}\sum_{n=0}^{\infty}\frac{eB}{(2\pi)^2}\int dp_{z}\bigg(E_{i,c}(p_{z},n,S_{z})+T\:\text{ln}(1\pm e^{-(E_{i,c}-\mu_i)/T})\bigg); e_i\neq 0
\label{idprechrg}
\end{equation}

\begin{equation}
P_{n}^{\text{id}}=\pm\sum_{i}\sum_{S_z}\int\frac{d^3p}{(2\pi)^3}\:\bigg(E_{i,n}+T\:\text{log}\bigg[1\pm \text{exp}\bigg(-\frac{(E_{i,n}-\mu_i)}{T}\bigg)\bigg]\bigg); e_i=0
\label{idpreneutral}
\end{equation}

Note that  the thermal part of the pressure is naturally convergent in the $UV$ limit but the vacuum part is divergent and needs to be regularized. 

\section{Regularization of vacuum pressure}
\label{secIII}

The vacuum pressure is divergent and it needs to be properly regularized first.  It has been recently shown that the appropriate regularization scheme is necessary to avoid certain unphysical results. For instance, some studies have found the oscillations in the magnetization while others have found imaginary meson masses~\cite{Fayazbakhsh:2013cha}. Both of these findings are unphysical and it can be attributed to an inappropriate regularization choices. These unphysical results arise especially in case of magnetic field dependent regularization schemes. Thus, it is absolutely important that magnetic field dependent and independent parts are separated clearly through appropriate regularization scheme. Magnetic field independent regularization (MFIR) has recently been introduced to achieve this goal~\cite{Ebert:2003yk,Menezes:2008qt}. We shall obtain the regularized vacuum  pressure for spin $\frac{1}{2}$ particles using MFIR method. Spin zero and spin one cases can be discussed in a  similar  manner(see \ref{vacpressure} for renormalized vacuum pressure expressions).

The vacuum part of the  pressure for a charged spin-$\frac{1}{2}$ particle in magnetic field  is 

\begin{equation}
	P_{\text{vac}}(S=1/2,B)=\sum_{n=0}^{\infty}g_{n}\frac{eB}{2\pi}\int_{-\infty}^{\infty}\frac{dp_z}{2\pi}E_{p,n}(B)
	\label{vacphalf}
\end{equation}
where $g_{n}=2-\delta_{n0}$ is the degeneracy of n$^{\text{th}}$ Landau level. Now adding and subtracting lowest Landau level contribution (i.e. $n=0$) from the above equation we get
\begin{equation}
P_{\text{vac}}(S=1/2,B)=\sum_{n=0}^{\infty}2\frac{eB}{2\pi}\int_{-\infty}^{\infty}\frac{dp_z}{2\pi}\bigg(E_{p,n}(B)-\frac{E_{p,0}(B)}{2}\bigg)
\label{vacphalf1}
\end{equation}

 We regularize the divergence using dimensional regularization~\cite{Peskin:1995}. In $d-\epsilon$ dimension Eq. (\ref{vacphalf1}) can be written as
 \begin{equation}
 P_{\text{vac}}(S=1/2,B)=\sum_{n=0}^{\infty}\frac{eB}{\pi}\mu^{\epsilon}\int_{-\infty}^{\infty}\frac{d^{1-\epsilon}p_z}{(2\pi)^{1-\epsilon}}\bigg(\sqrt{p_z^2+m^2-2eBn}-\sqrt{p_z^2+m^2}\bigg)
 \label{vacphalf2}
 \end{equation}
 where $\mu$ fixes the dimension of the above expression to one.  The integration can be done using standard $d-$dimensional formula (see \ref{formulae}). Integration of the first term in  Eq.(\ref{vacphalf2}) gives
 \begin{equation}
I_1=\sum_{n=0}^{\infty}\frac{eB}{\pi}\mu^{\epsilon}\int_{-\infty}^{\infty}\frac{d^{1-\epsilon}p_z}{(2\pi)^{1-\epsilon}}(p_z^2+m^2-2eBn)^{\frac{1}{2}}=-\frac{(eB)^2}{2\pi^2}\bigg(\frac{2eB}{4\pi\mu}\bigg)^{-\frac{\epsilon}{2}}\Gamma\bigg(-1+\frac{\epsilon}{2}\bigg)\zeta\bigg(-1+\frac{\epsilon}{2},x\bigg)
\label{I1}
\end{equation}
where $x\equiv\frac{m^2}{2eB}$. The Landau infinite sum has been expressed in terms of  Riemann-Hurwitz $\zeta-$function (see Eq. (\ref{RHdef})). Using the expansion of $\Gamma$-function (see Eq.(\ref{gammaexp1})) and the expansion of $\zeta$-function (see Eq.(\ref{zetaexp})), Eq.(\ref{I1}) can be written as
 \begin{equation}
 I_1=-\frac{(eB)^2}{2\pi^2}\bigg(-\frac{2}{\epsilon}+\gamma-1+\text{ln}\bigg(\frac{2eB}{4\pi\mu^2}\bigg)\bigg)\bigg(-\frac{1}{12}-\frac{x^2}{2}+\frac{x}{2}+\frac{\epsilon}{2}\zeta^{'}(-1,x)+\mathcal{O}(\epsilon^2)\bigg)
 \end{equation}
 
 Integration of the second term in  Eq.(\ref{vacphalf2}) can be simplified in similar manner. We obtain
 \begin{eqnarray}
 I_2&=&\sum_{n=0}^{\infty}\frac{eB}{\pi}\mu^{\epsilon}\int_{-\infty}^{\infty}\frac{d^{1-\epsilon}p_z}{(2\pi)^{1-\epsilon}}(p_z^2+m^2)^{\frac{1}{2}}\nonumber\\
     &=&\frac{(eB)^2}{2\pi^2}\bigg(-\frac{x}{\epsilon}-\frac{(1-\gamma)}{2}x+\frac{x}{2}\text{ln}\bigg(\frac{2eB}{4\pi\mu^2}\bigg)+\frac{x}{2}\text{ln}(x)\bigg)
 \end{eqnarray}
 
 Thus the vacuum pressure in presence of magnetic field becomes
 \begin{eqnarray}
 P_{\text{vac}}(S=1/2,B)&=&\frac{(eB)^2}{2\pi^2}\bigg(\zeta^{'}(-1,x)-\frac{2}{12\epsilon}-\frac{(1-\gamma)}{12}-\frac{x^2}{\epsilon}-\frac{(1-\gamma)}{2}x^2\nonumber\\
 &+&\frac{x}{2}\text{ln}(x)+\frac{x^2}{2}\text{ln}\bigg(\frac{2eB}{4\pi\mu^2}\bigg)+\frac{1}{12}\text{ln}\bigg(\frac{2eB}{4\pi\mu^2}\bigg)\bigg)
                                         \label{vacphalfB}
 \end{eqnarray}
 
 Above expression is still divergent. So we add and subtract $B=0$ contribution from it. This zero field vacuum pressure in $d=3-\epsilon$ dimension is
 \begin{eqnarray}
P_{\text{vac}}(S=1/2,B=0)&=& 2\mu^{\epsilon}\int\frac{d^{3-\epsilon}p}{(2\pi)^{3-\epsilon}}\:(p^2+m^2)^{\frac{1}{2}}\nonumber\\
                                       &=&\frac{(eB)^2}{2\pi^2}\bigg(\frac{2eB}{4\pi\mu^2}\bigg)^{-\frac{\epsilon}{2}}\Gamma\bigg(-2+\frac{\epsilon}{2}\bigg)x^{2-\frac{\epsilon}{2}}
 \end{eqnarray}
 
 Above equation can be further simplified to
 \begin{equation}
 P_{\text{vac}}(S=1/2,B=0)=-\frac{(eB)^2}{2\pi^2}x^2\bigg(\frac{1}{\epsilon}+\frac{3}{4}-\frac{\gamma}{2}-\frac{1}{2}\text{ln}\bigg(\frac{2eB}{4\pi\mu^2}\bigg)-\frac{1}{2}\text{ln}(x)\bigg)
 \label{vachalfB0}
 \end{equation}
 where we have used the $\Gamma$-function expansion (see Eq. \ref{gammaexp2}).

 Adding and subtracting (\ref{vachalfB0}) from (\ref{vacphalfB}) we get the regularized  pressure with vacuum part and magnetic field dependent part separated as
 \begin{eqnarray}
 P_{\text{vac}}(S=1/2,B)&=&P_{\text{vac}}(1/2,B=0)+\Delta P_{\text{vac}}(1/2,B)
 \label{pvacB}
 \end{eqnarray}
 where
 \begin{eqnarray}
 \Delta P_{\text{vac}}(S=1/2,B)&=&\frac{(eB)^2}{2\pi^2}\bigg(-\frac{2}{12\epsilon}+\frac{\gamma}{12}+\frac{1}{12}\text{ln}\bigg(\frac{m^2}{4\pi\mu^2}\bigg)+\frac{x}{2}\text{ln}(x)\nonumber\\
 &-&\frac{x^2}{2}\text{ln}(x)+\frac{x^2}{4}-\frac{\text{ln}(x)+1}{12}+\zeta^{'}(-1,x)\bigg)
 \label{delpvacB}
 \end{eqnarray}

 The field contribution given by (\ref{delpvacB}) is still divergent due to presence of pure magnetic field dependent term $\propto \frac{B^2}{\epsilon}$~\cite{Schwinger:1951nm,Elmfors:1993bm,Andersen:2011ip}. We cancel this divergence by redefining field dependent pressure contribution by including magnetic field contribution in it as
 \begin{equation}
 \Delta P_{\text{vac}}^{r}=\Delta P_{\text{vac}}(B)-\frac{B^2}{2}
 \end{equation}
 
 The divergences are absorbed into the renormalization of the electric charge and magnetic field strength~\cite{Endrodi:2013cs},
 \begin{equation}
 B^2=Z_{e}B_r^2; \hspace{0.5cm} e^2=Z_e^{-1}e_r^2; \hspace{0.5cm}  e_rB_r=eB
 \end{equation}
 where the electric charge renormalization constant is
 \begin{equation}
 Z_{e}\bigg(S=\frac{1}{2}\bigg)=1+\frac{1}{2}e_r^2\bigg(-\frac{2}{12\epsilon}+\frac{\gamma}{12}+\frac{1}{12}\text{ln}\bigg(\frac{M_{*}}{4\pi\mu^2}\bigg)\bigg)
 \end{equation}
 
 Here we fix $M_{*}=m$, i.e to the physical mass of the particle. Thus the renormalized field dependent pressure (without pure magnetic field contribution) is
 \begin{equation}
\Delta P_{\text{vac}}^r(S=1/2,B)=\frac{(eB)^2}{2\pi^2}\bigg(\zeta^{'}(-1,x)+\frac{x}{2}\text{ln}(x)
-\frac{x^2}{2}\text{ln}(x)+\frac{x^2}{4}-\frac{\text{ln}(x)+1}{12}\bigg)
\label{delpvacBfin}
 \end{equation}
 
 The renormalized $B$ dependent pressure for spin zero and spin one can be obtained using similar method. These terms play very crucial role in determining the magnetization of the hadronic matter below $T_c$.  Note that the vacuum pressure depends on the quantum numbers like spin, mass, charge etc. Thus the total vacuum  pressure of hadron gas is evaluated by adding the vacuum pressure of all the hadronic species included in HRG model.

\section{Interacting Hadron resonance gas model in magnetic field}
\label{secIV}
   The hadron resonance gas model, defined by Eqs.(\ref{idprechrg}) and (\ref{idpreneutral}), corresponds to non-interacting gas of hadrons. One can extend this ideal HRG model by taking in to account the repulsive interactions between hadrons via Van der Waals (VDW) excluded-volume correction to the partition function.  In the thermodynamically consistent excluded volume formulation one can obtain the transcendental equation for the (thermal part of) pressure as~\cite{Rischke:1991ke}

\begin{equation}
P^{EV}(T,\mu,B)=P^{\text{id}}(T,\tilde{\mu},B),
\label{prexcl}
\end{equation}
where $\tilde{\mu}=\mu-vP^{EV}(T,\mu,B)$ is an effective chemical potential with $ v $ as the parameter corresponding to proper volume of the particle (excluded-volume parameter) 
which is actually the second virial coefficient. Here $P_{id}$ is the ideal gas pressure given by Eqs. (\ref{idprechrg}) and (\ref{idpreneutral}) without vacuum part. For a particle of hard core radius $r_h$,  $ v=\frac{16}{3}\pi r_{h}^3 $ the excluded volume pressure can be obtained by solving Eq.(\ref{prexcl}) self consistently for given $T,\mu$ and $B$. 

 The only free parameter in our calculation is hadron hardcore radius, $r_h$.  It has been found in 
detailed studies~\cite{zeeb,bugaev,vovchenko1} that thermal models with the multi-component hard-core radii  describe the hadron yield ratios with improved 
quality  if different hard-core radii are chosen for different hadrons. One such model~\cite{bugaev} has also been shown to describe the Strangeness Horn behaviour.   
Here we take the sizes of hadrons as prescribed in Ref.~\cite{bugaev} i.e. $r_\pi$ (pion radius) $= 0$ fm, $r_K$ (kaon radius) $= 0.35$ fm, $r_m$ (all other meson radii) 
$= 0.3$ fm  and $r_b$ (baryon radii) $= 0.5$ fm.

\section{Results and discussion}
\label{secV}

Let us now  discuss the results we have obtained. We would like to mention at this stage that the results have been obtained around the point $\mu_Q=0$ and $\mu_S=0$ 
and  both zero and finite $\mu_B$ cases have been explored.

Table \ref{tab:hadrons} shows all  the hadrons and resonances included in the HRG description. For the stability reasons, as discussed in Ref.~\cite{Endrodi:2013cs}, we include only those hadrons having spin $S<\frac{3}{2}$. 

 \begin{table}[h!]
	\centering
	\begin{tabular}{|l|c|c|c|c||l|c|c|c|c|}
\hline
hadron & $m (\textmd{GeV})$ & $|e|$ & Spin & $\textmd{deg.}$ & hadron & $m (\textmd{GeV})$ & $|e|$ & Spin & $\textmd{deg.}$ \\ 
\hline
\hline
$\pi^{\pm}$ & 0.135 & 1 & 0 & 2  & $p$ & 0.938 & 1 & 1/2 & 2  \\
$\pi^0$ & 0.135 & 0 & 0 & 1 & $n$ & 0.938 & 0 & 1/2 & 2  \\
$K^{\pm}$ & 0.495 & 1 & 0 & 2 & $\eta'$ & 0.958 & 0 & 0 & 1  \\
$K^0$ & 0.495 & 0 & 0 & 2 & $f_0$ & 0.980 & 0 & 0 & 1  \\
$\eta$ & 0.548 & 0 & 0 & 1 & $a_0$ & 0.980 & 0 & 1 & 1  \\ 
$\rho^{\pm}$ & 0.776 & 1 & 1 & 2 & $\phi$ & 1.020 & 0 & 1 & 1  \\
$\rho$ & 0.776 & 0 & 1 & 1 & $\Lambda$ & 1.116 & 0 & 1/2 & 1 \\  
$\omega$ & 0.782 & 0 & 1 & 1 & $h_1$ & 1.170 & 0 & 1 & 1 \\ 
$K_{*}^{\pm}$ & 0.892 & 1 & 1 & 2 & $\Sigma^\pm$ & 1.189 & 1 & 1/2 & 2 \\ 
$K_*^0$ & 0.892 & 0 & 1 & 2 &  $\Sigma^0$ & 1.189 & 0 & 1/2 & 1 \\ 
\hline
\end{tabular}

	\caption{\label{tab:hadrons}(Color Online)  Hadrons and resonances  included in the hadron resonance gas model. Particle data is taken from Ref.~\cite{Beringer:1900zz}.
	}
\end{table}

\begin{figure}[t]
\vspace{-0.4cm}
\begin{center}
\begin{tabular}{c c}
 \includegraphics[width=7cm,height=7cm]{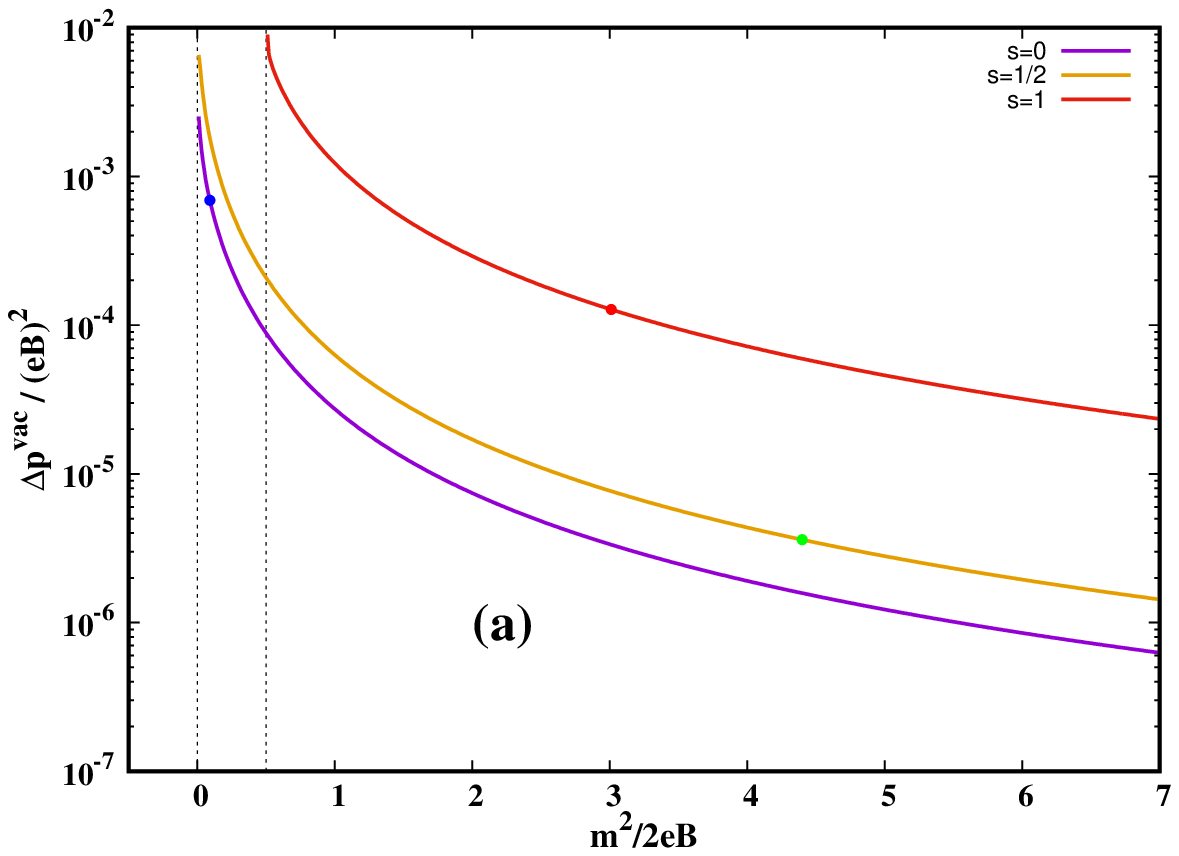}&
  \includegraphics[width=7cm,height=7cm]{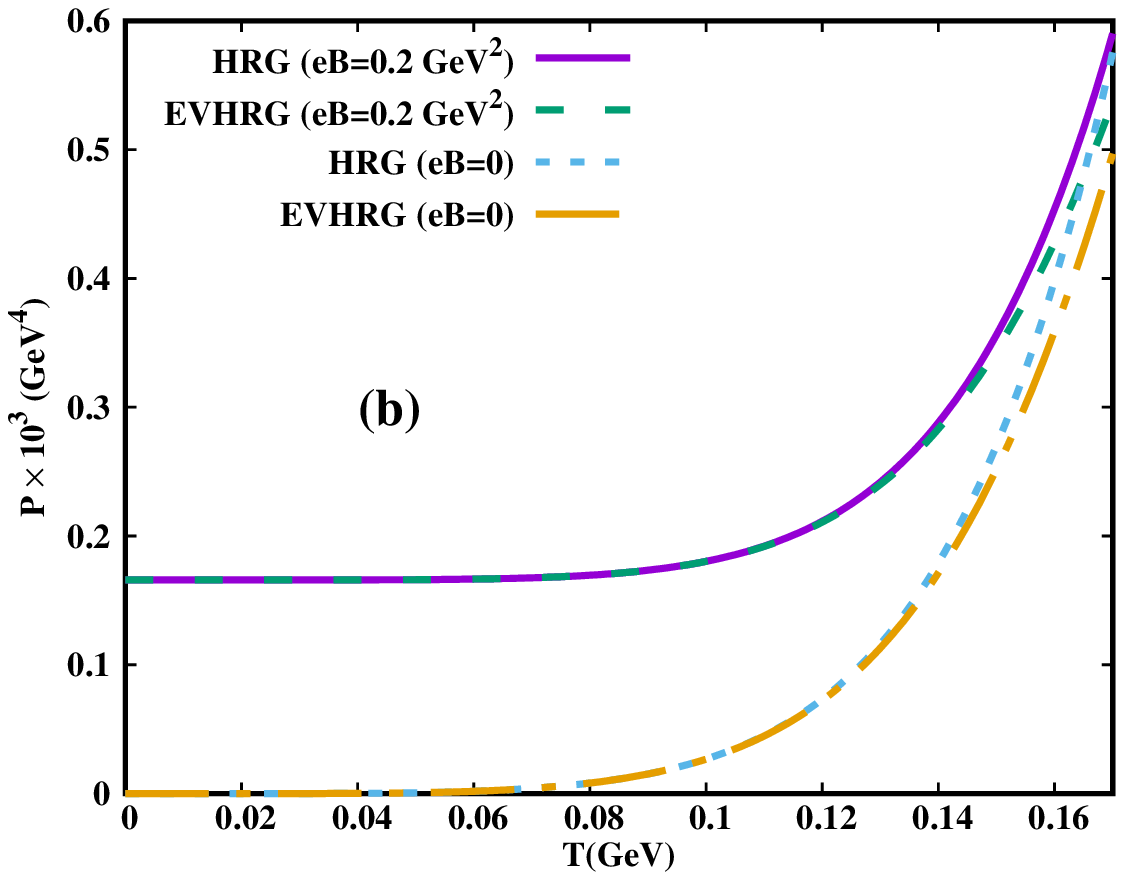}
  \end{tabular}
  \caption{(Color Online) Left panel shows vacuum pressure of charged particles computed using MFIR scheme plotted against dimensionless variable $x=\frac{m^2}{2eB}$.
   Right panel shows the total pressure as a function of temperature in presence of magnetic field for $\mu_B=0$.   } 
\label{pressure}
  \end{center}
 \end{figure}

Fig.\ref{pressure}(a) shows the variation of the vacuum  pressure as a function of the dimensionless variable $x=\frac{m^2}{2eB}$. For spin-0, spin $\frac{1}{2}$ and spin-1 hadrons the vacuum pressure is positive for wide range of magnetic fields and hence we can safely assume that the HRG description is valid.  Fig.\ref{pressure}(b) shows the variation of  total pressure with temperature for HRG and EVHRG models both in presence and in absence of magnetic field at zero $\mu_B$. Note that the pressure in presence of magnetic field has magnetic field dependent vacuum contribution. Such contribution is zero if $eB=0$. Hence the HRG pressure without magnetic field vanishs at $T=0$ GeV while it is non-zero at finite magnetic field. This is the reason we have plotted the pressure instead of scaled pressure ($\frac{P}{T^4}$).  We further note that the pressure increases with temperature in all cases considered, as the contribution to pressure of a 
particular species, with mass $m$,  is proportional to the Boltzmann factor $e^{-m/T}$. It is necessary here to point out that, for HRG, the thermal part of the pressure in presence of magnetic field is smaller than that for $eB=0$ case. However, it is the other way round for the  vacuum part of the pressure.  This can be accounted  by comparing  the dispersion relations in presence of  magnetic field (Eq.(\ref{dispmagn})) and in the absence of magnetic field (Eq.(\ref{disp})). The effective mass of the charged particle in presence of magnetic field is $m_{eff}^2=m^2+2eB(\frac{1}{2}-S)$. Thus the mass of a spin-0 particle increases  whereas that of spin-1 particle decreases if $eB\neq 0$. For spin-$\frac{1}{2}$ particles mass remains unchanged. Since the particles with spin-1 are heavier (lightest spin-1 particle $\rho$ weighs 0.776 GeV which is more than five times heavier than pion), their contributions to pressure  are much smaller compared to spin-0  particles and show up only at higher temperature. Since pions dominate the hadronic matter at low temperature, and since their effective mass increases in presence of magnetic field, the thermal part of the pressure, for HRG model, in presence of magnetic field is smaller than that for without magnetic field.  

Fig.\ref{pressure}(b) further shows the effect of repulsive interaction on the pressure. If we include the repulsive interactions through excluded volume correction to ideal gas partition function then  the pressure is suppressed for both $eB=0$ and $eB\neq0$ cases. With finite size of hadrons the available free space for hadrons decreases with increasing temperature. This decreases the number density and hence the pressure of hadrons compared to free HRG case. From Fig.\ref{pressure} (b) it is seen that  pressures for HRG and EVHRG models are almost identical up to $T\sim0.1 $ GeV  for $eB=0$ case and up to $T\sim0.12$ GeV  for $eB=0.2 \text{GeV}^2$ case. Above these temperatures there is a notable decrease in  pressure for EVHRG case.

  \begin{figure}[t]
\vspace{-0.4cm}
\begin{center}
\begin{tabular}{c c}
\includegraphics[width=7cm,height=7cm]{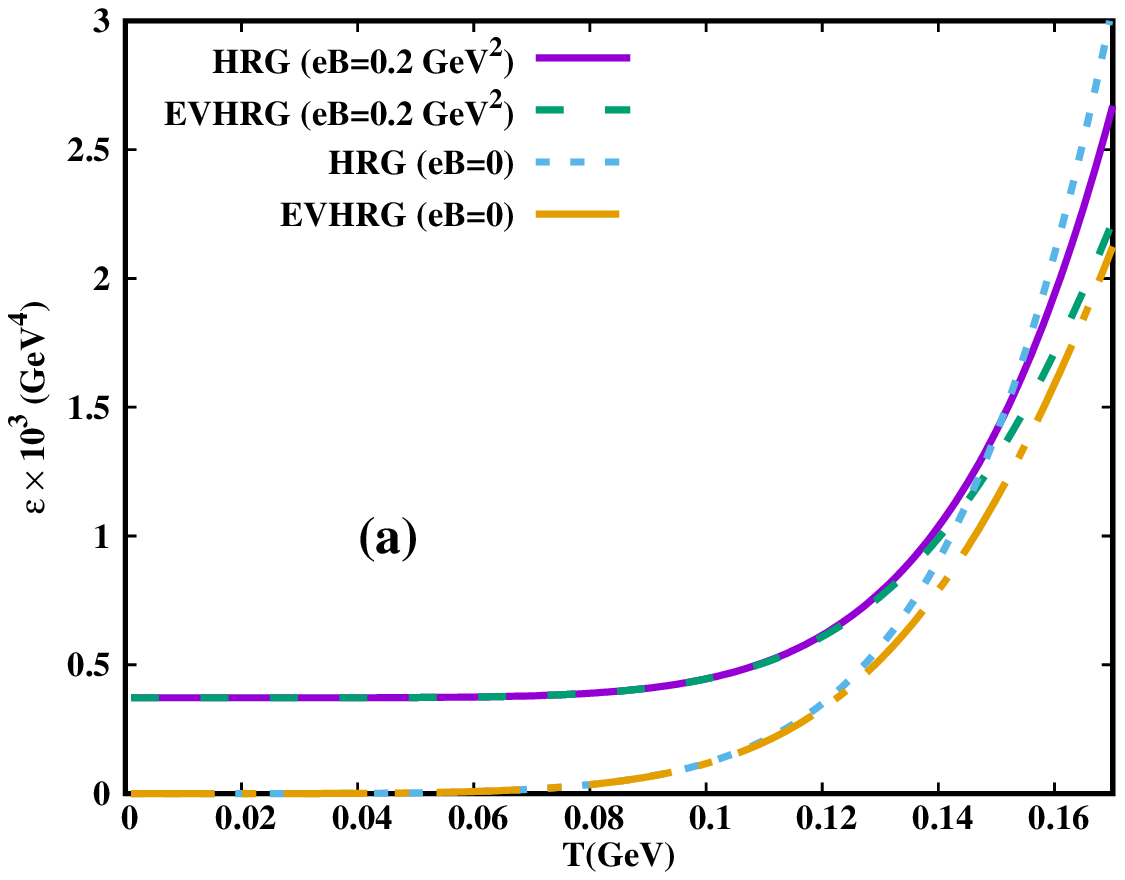}&
  \includegraphics[width=7cm,height=7cm]{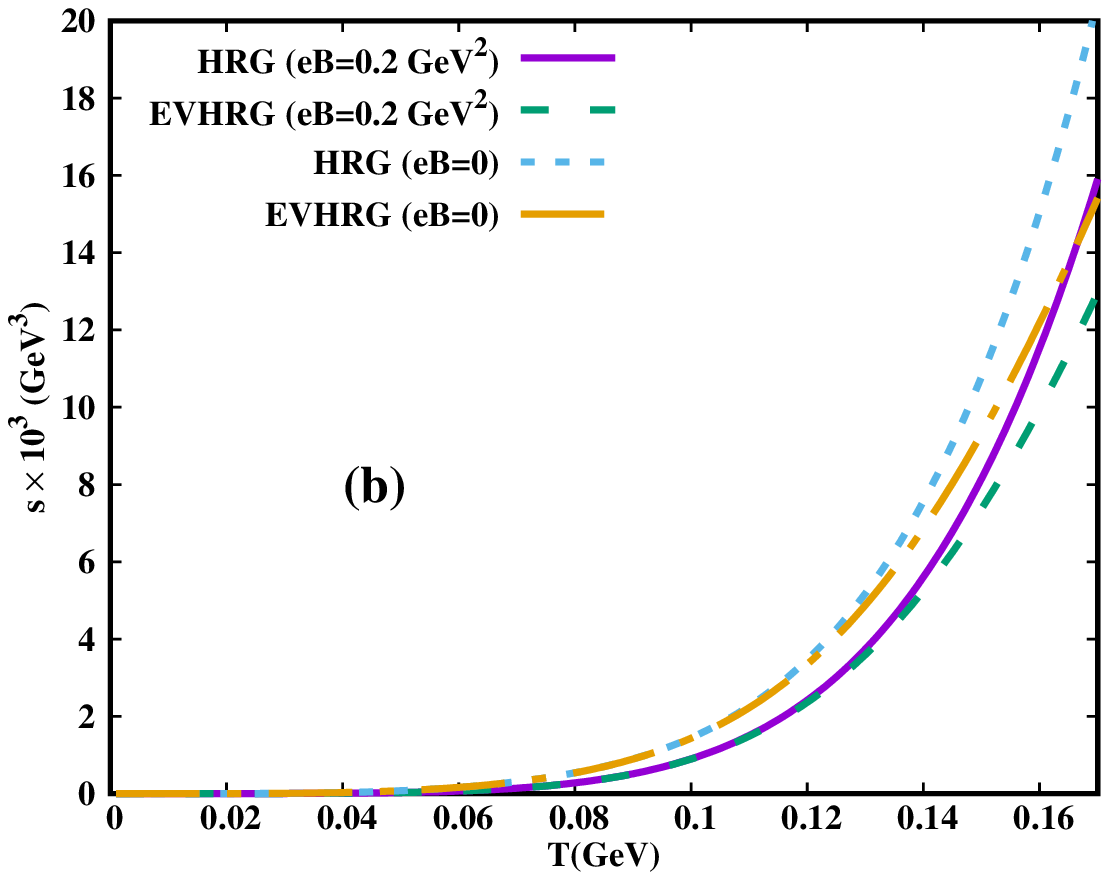}
  \end{tabular}
  \caption{(Color Online) Energy density (left panel) and  entropy density (right panel) calculated in HRG and EVHRG models with and without  magnetic field for $\mu_B=0$. } 
\label{energy}
  \end{center}
 \end{figure}

 Fig.\ref{energy} shows the effect of magnetic field on the energy density and entropy density for both HRG and EVHRG models at zero $\mu_B$. The effect of magnetic field as well as repulsive interactions on energy density  is similar to that of pressure. The thermal part of the energy density in presence of magnetic field is smaller than that of $eB=0$ case. The repulsive interactions further suppress the energy density.  The effect of magnetic field as well as the repulsive interactions on the entropy density is quite interesting. In HRG, the entropy density rises rapidly as the temperature increases due to copious production of hadrons. As we have discussed above, the effective mass of a spin-0 particle increases in presence of magnetic field. Thus their thermal production is accordingly suppressed by Boltzmann factor $e^{-m_{eff}/T}$ and the entropy production is suppressed in presence of magnetic field. The  repulsive interaction also suppresses the entropy productions since the finite size of hadrons suppresses the number density at high temperature.

   \begin{figure}[b]
\vspace{-0.4cm}
\begin{center}
\begin{tabular}{c c}
  \includegraphics[width=7cm,height=7cm]{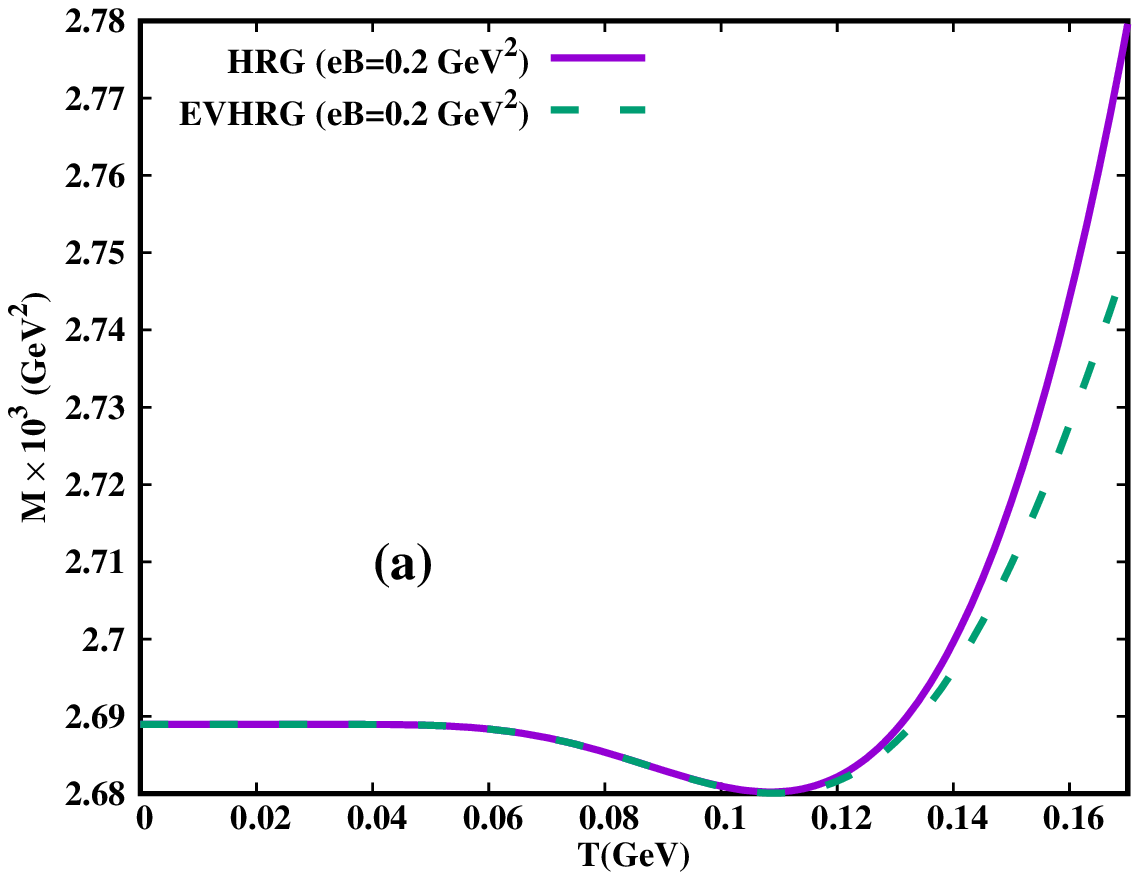}&
  \includegraphics[width=7cm,height=7cm]{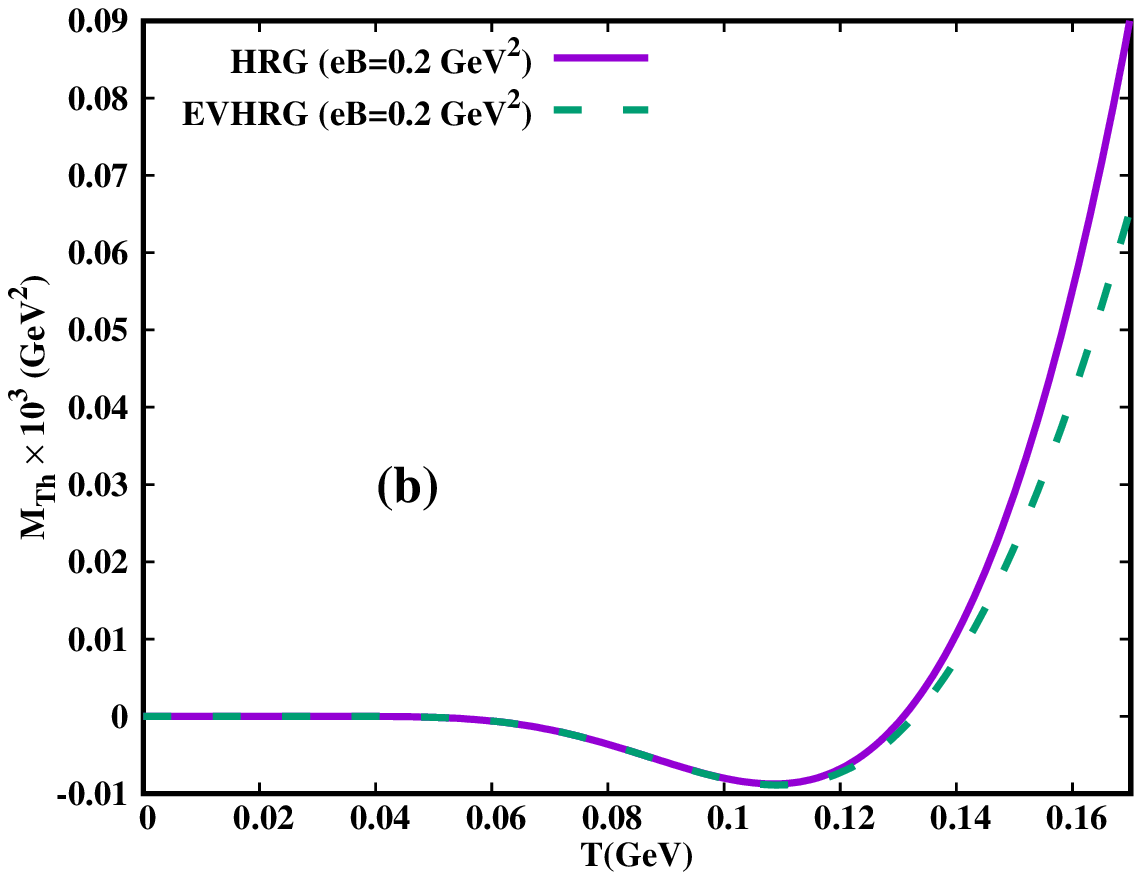}
  \end{tabular}
  \caption{(Color Online) Magnetization of hadronic matter estimated using HRG and EVHRG in presence of magnetic field. Left panel shows the total (vacuum+thermal)
  magnetization and the right panel shows only the thermal part of magnetization. Calculations are performed at  $\mu_B=0$} 
\label{magnetization}
  \end{center}
 \end{figure}
 
In Fig.\ref{magnetization} we have plotted magnetization as a function of temperature for zero $\mu_B$. Fig.\ref{magnetization}(a) shows the behaviour of total magnetization.  The magnetization is positive indicating that the hadronic matter is paramagnetic. Fig.\ref{magnetization}(b) shows only thermal contribution to the magnetization. At very low temperature the thermal part of magnetization is practically zero due to the absence of charged hadrons.  We note that the pions, which are lightest hadronic species, are thermally excited at $T\sim 0.060$ GeV. Being scalar bosons their magnetization is negative (diamagnetic) and hence magnetization decreases with increase in temperature. It becomes positive only when lightest spin-1 particle, $\rho$-meson, populates the hadronic matter and gives positive contribution to hadronic matter. Thereafter, the magnetization rises rapidly as spin-$\frac{1}{2}$ particles also start to make positive contribution to the magnetization. It is interesting to note that even though the thermal part of magnetization becomes negative for certain temperature range the total magnetization including vacuum part is always positive. It turns out that the sign of magnetization of low temperature hadronic matter is a fundamental characteristic of thermal QCD vacuum.

\begin{figure}[t]
\vspace{-0.4cm}
\begin{center}
\begin{tabular}{c c}
 \includegraphics[scale=0.6]{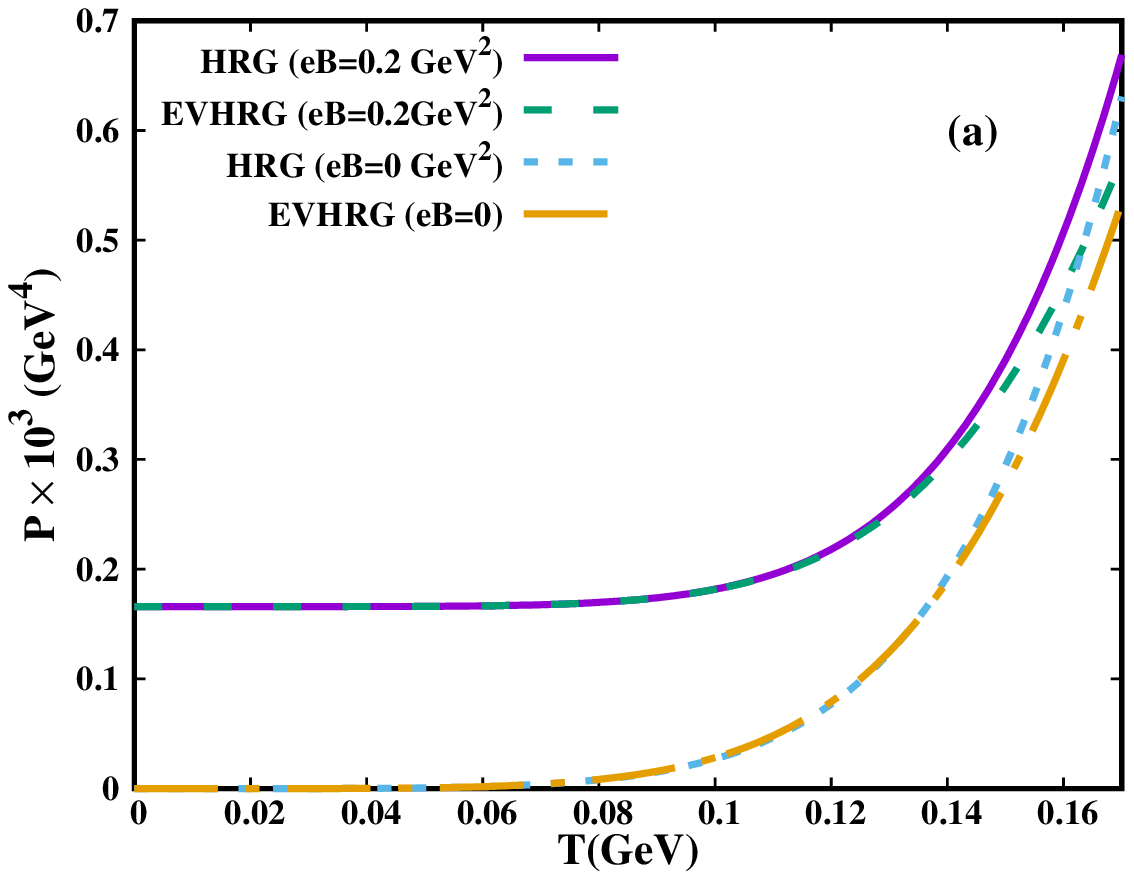}
  \includegraphics[scale=0.6]{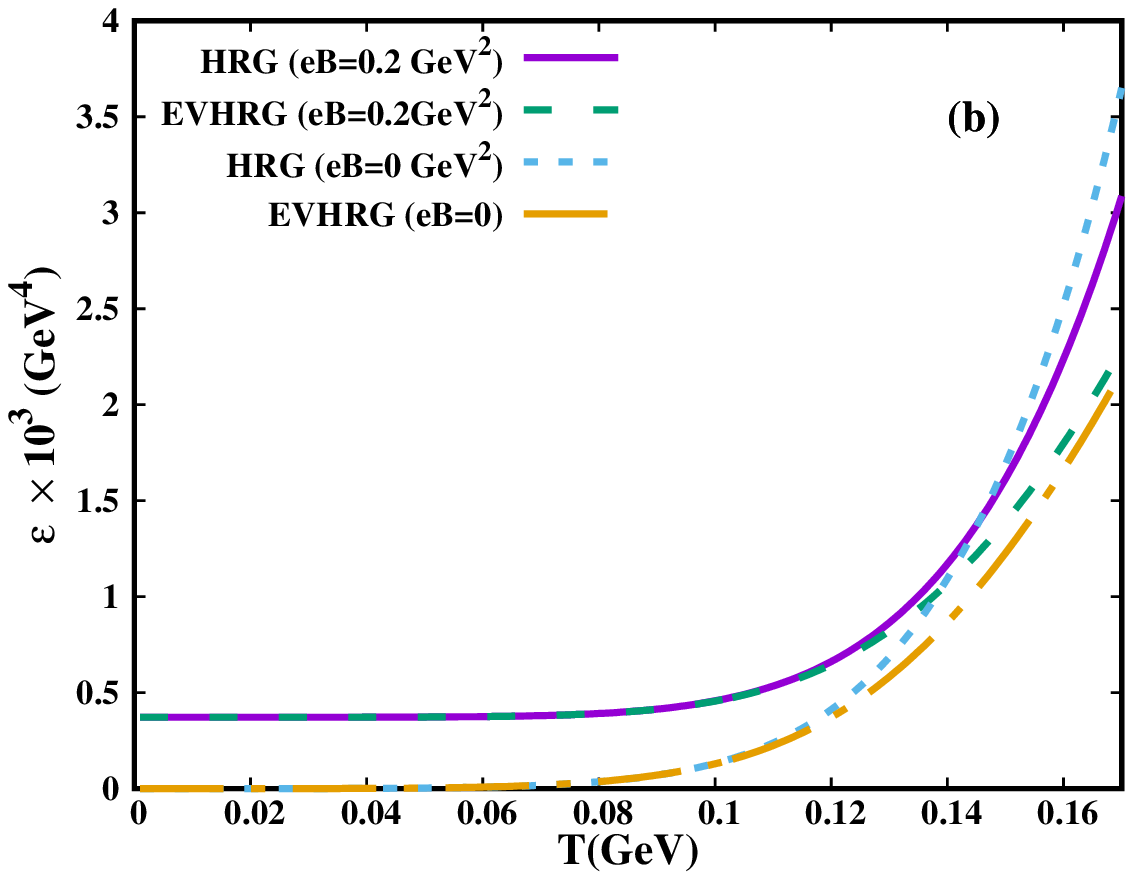} \\ 
  \includegraphics[scale=0.6]{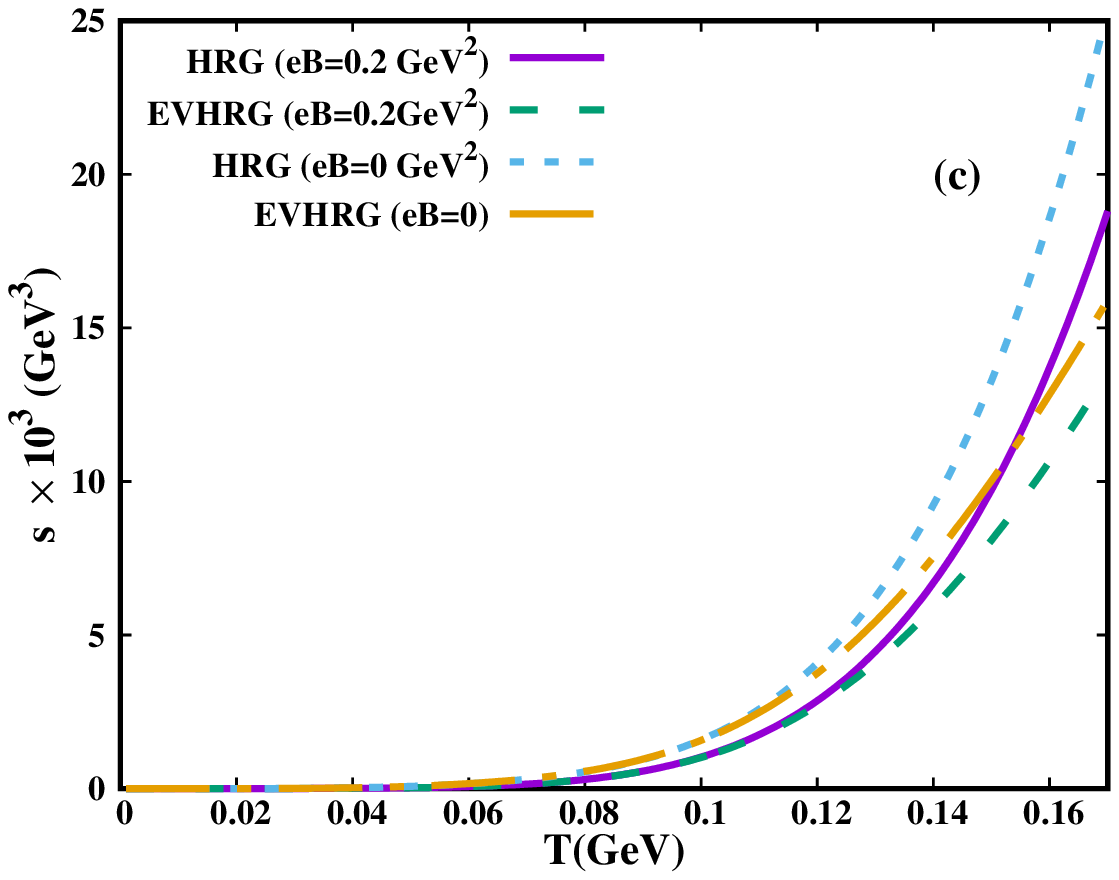}
  \includegraphics[scale=0.6]{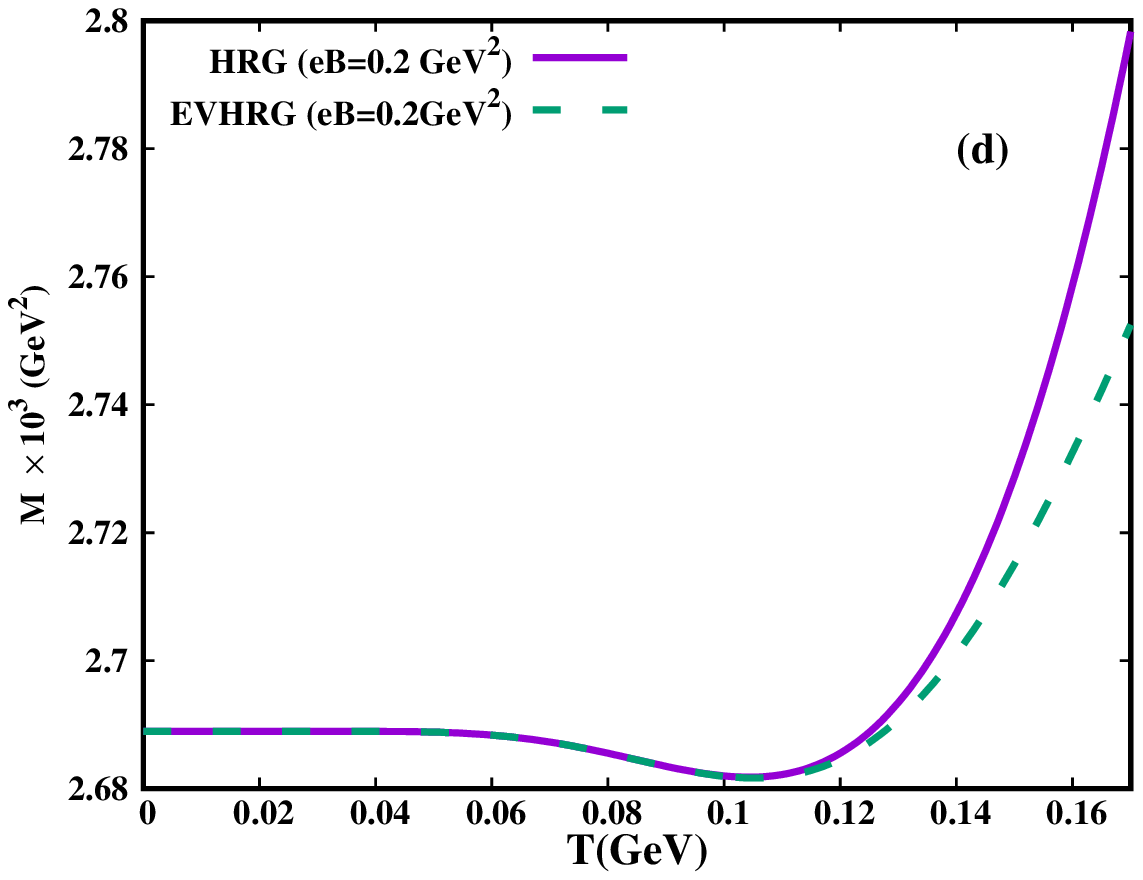} 
  \end{tabular}
  \caption{(Color Online) Thermodynamic variables at $\mu_B=300 MeV$. The results are for  HRG and EVHRG models with and without  magnetic field.
  a) Pressure, b) Energy Density, c) Entropy density and d) Magnetization. } 
\label{figmu}
  \end{center}
 \end{figure}
  In Fig.\ref{figmu} we have plotted the pressure, energy density, entropy density and magnetization at finite $\mu_B$. We choose  $\mu_B$ to be 300 MeV 
  which is close to CEP. In all the plots of this figure we see that  the qualitative behaviour of all the thermodynamic quantities are similar to that obtained 
  for the $\mu_B=0$ case.    In Fig.\ref{figmu}(a) the pressure has been plotted. A comparison with Fig~\ref{pressure}(b) shows that, at high temperatures, for $eB=0$,  
  the pressure at finite $\mu_B$ is more than that at zero $\mu_B$. This increase, for HRG, is about $5\%$ whereas for EVHRG the pressure does not have any noticeable 
  change. For $eB=0.2 GeV^2$ the  increase in pressure, at finite $\mu_B$, for HRG and EVHRG, are  $16\%$ 
  and $9\%$ respectively. Once we compare energy density (Fig.\ref{energy}(a) and Fig.\ref{figmu}(b)), at zero and finite $\mu_B$,   a similar situation is revealed.  However, 
  the magnitude of change, in the energy density, at finite $\mu_B$, is much more compared to that in pressure. 
  For $eB=0$ the energy density increases at finite $\mu_B$ compared to zero $\mu_B$ case, for HRG, by about $25\%$ whereas for EVHRG it does not change. For $eB=0.2 GeV^2$ the corresponding changes are  $19\%$ 
  and $8\%$ respectively. For entropy density, the changes for HRG, are similar as to that of energy density whereas for EVHRG there is almost no change. We do not 
  find any appreciable effect of finite $\mu_B$ on magnetization. A very interesting fact emerges from the above discussion. We find that the effect of finite $\mu_B$ 
  is much less, at times negligible, for EVHRG. This is because of the fact that as $\mu_B$ increases the suppression due to excluded volume also increases. 
  
  \begin{figure}[b]
\vspace{-0.4cm}
\begin{center}
\begin{tabular}{c c c}
 \includegraphics[width=5cm,height=5cm]{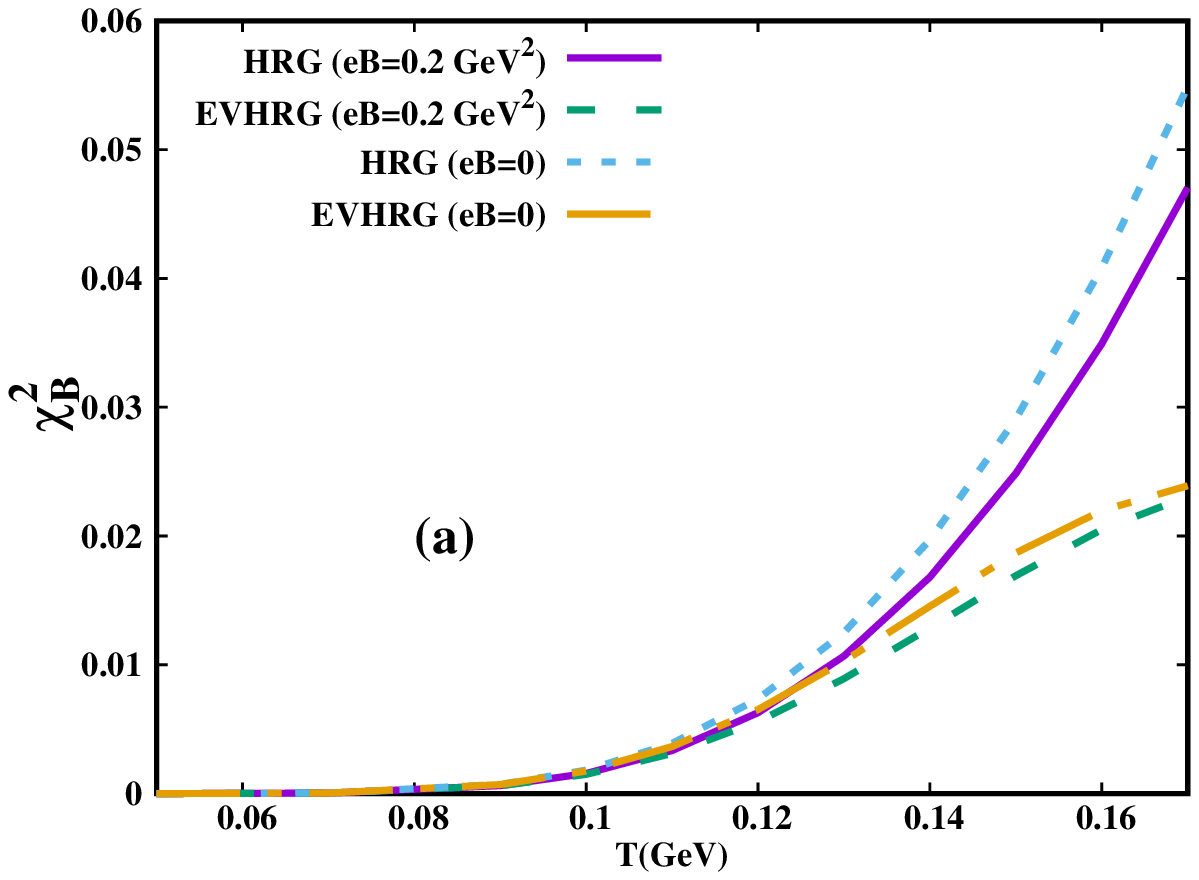}&
  \includegraphics[width=5cm,height=5cm]{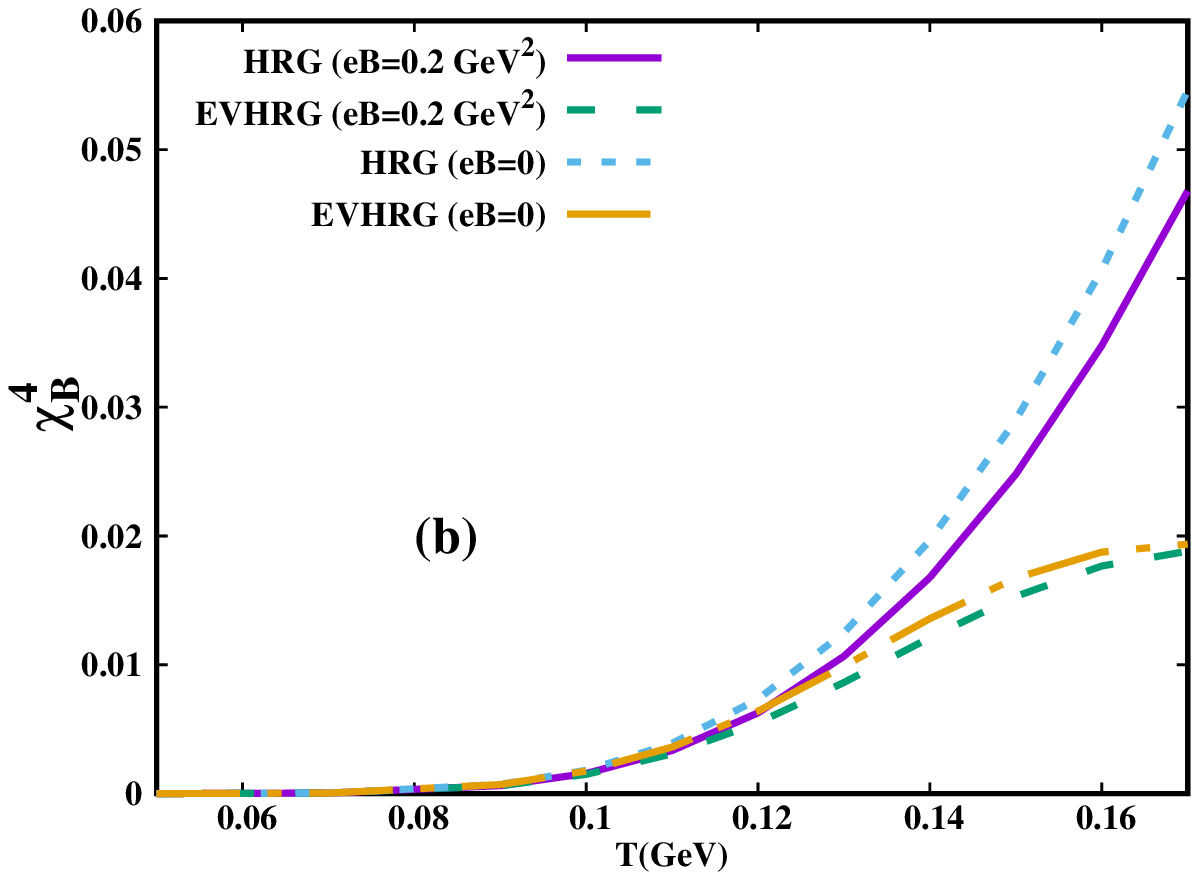}&
  \includegraphics[width=5cm,height=5cm]{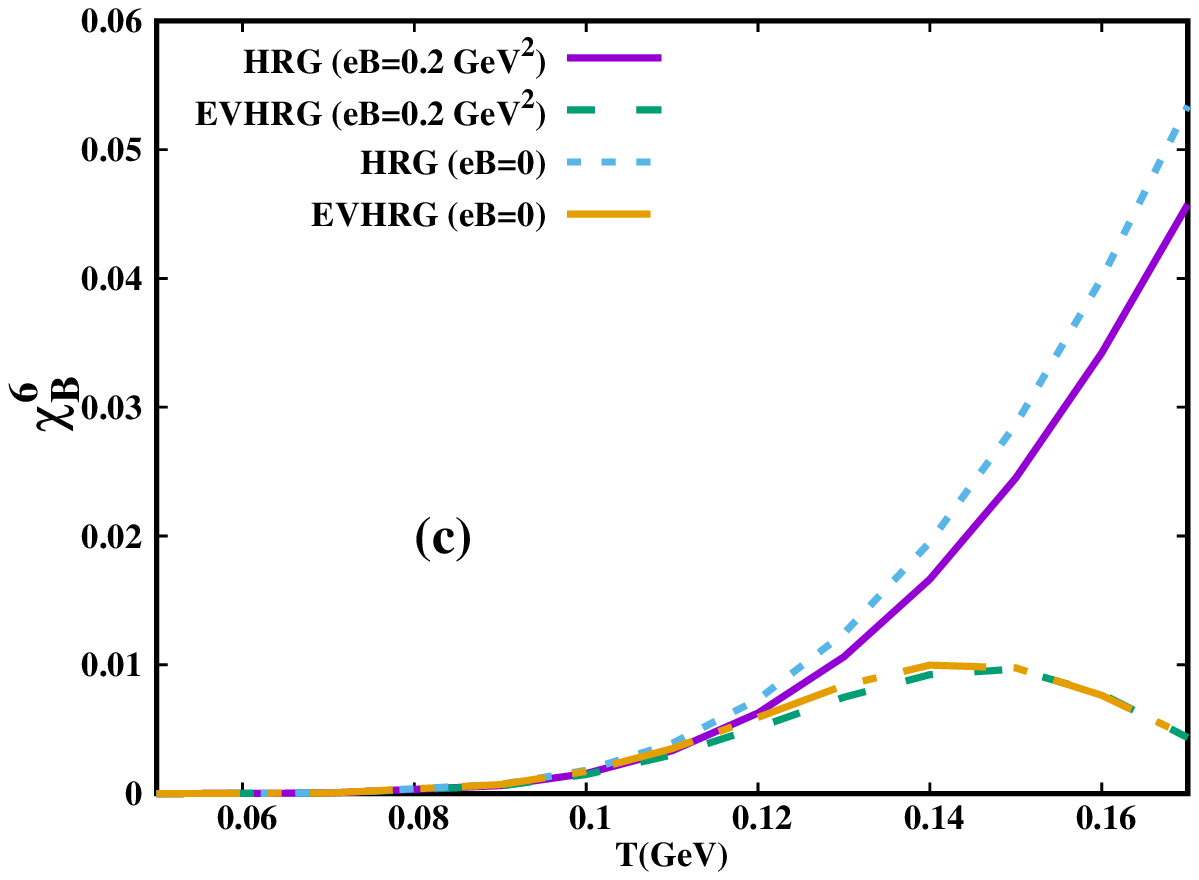}
  \end{tabular}
  \caption{(Color Online) Baryon susceptibilities of the hadronic matter estimated using HRG and EVHRG with and without magnetic field. } 
\label{chiB}
  \end{center}
 \end{figure}
 
Let us now discuss the fluctuation of conserved charges in presence of a magnetic field. The nth-order susceptibility is defined as 
 
 \begin{equation}
 \chi_{x}^{n}=\frac{1}{VT^3}\frac{\partial^n(\text{ln}Z)}{\partial (\frac{\mu_{x}}{T})^n}
 \end{equation}
 where $\mu_x$ is the chemical potential for conserved charge $x$. In this work we will take $x$ to be baryon number ($B$) and electric charge ($Q$).
 
 Fig.\ref{chiB} shows baryon number susceptibilities estimated within HRG and EVHRG models in presence of magnetic field. In case of HRG model 
 $\chi_B^{n}$s ($n=2,4,6$) increase rapidly at high temperature.  We note that the susceptibilities at a given temperature in presence of magnetic field is less than that for $eB=0$.  This is because the effect of magnetic field is to decrease the contribution to pressure from spin-$\frac{1}{2}$ charged particles as compared to the zero magnetic field case. 
 (We haven't considered any baryon with spin greater than $\frac{1}{2})$.  At low $T$, the dominating contribution to $\chi_{B}$ comes from nucleons, namely protons and neutrons which carry baryon number $1$. Note that the mesons, although they are predominant degrees of freedom at all temperatures, do not contribute to $\chi_{B}$ directly since they do not carry baryon number.   The probability that a baryon of mass $m$ populates at a given temperature $T$ is proportional to the Boltzmann factor $e^{-m/T}$. Thus as temperature increases other heavier baryons are thermally excited  and start to contribute to the pressure and hence to $\chi_{B}$. Here all the baryons have baryon number $\pm1$. For such a case there is almost no 
 difference in magnitude for  susceptibilities of different orders for $\chi_B$ in case of HRG model. We will explain this issue later. 
 
 The effects of repulsive interactions accounted through excluded volume corrections can also be seen in Fig.\ref{chiB}. One can  see that the values of $\chi_B^n$ with repulsive interactions are smaller as compared to those calculated in ideal HRG model. The effect of repulsive interaction is not significant at low temperatures since the population of various particles is small at low temperature and hence there is enough room for particles in the system. It is well known fact that the presence of repulsive interaction suppresses the pressure at high temperature as available space for additional number of hadrons and resonances decreases. This suppression is traded in to the $\chi_B^n$. Thus, the presence of repulsive interactions  affect the baryon number susceptibilities and, unlike the ideal HRG, this suppression effect becomes stronger for higher order $\chi_{B}^n$s i.e. $\chi_B^n$s of different orders are not of same magnitude when interaction is present. The $\chi_B^6$ initially increases with temperature and then decreases for both $eB=0$ and $eB=0.2 GeV^2$.

 \begin{figure}[h]
\vspace{-0.4cm}
\begin{center}
\begin{tabular}{c c c}
 \includegraphics[width=5cm,height=5cm]{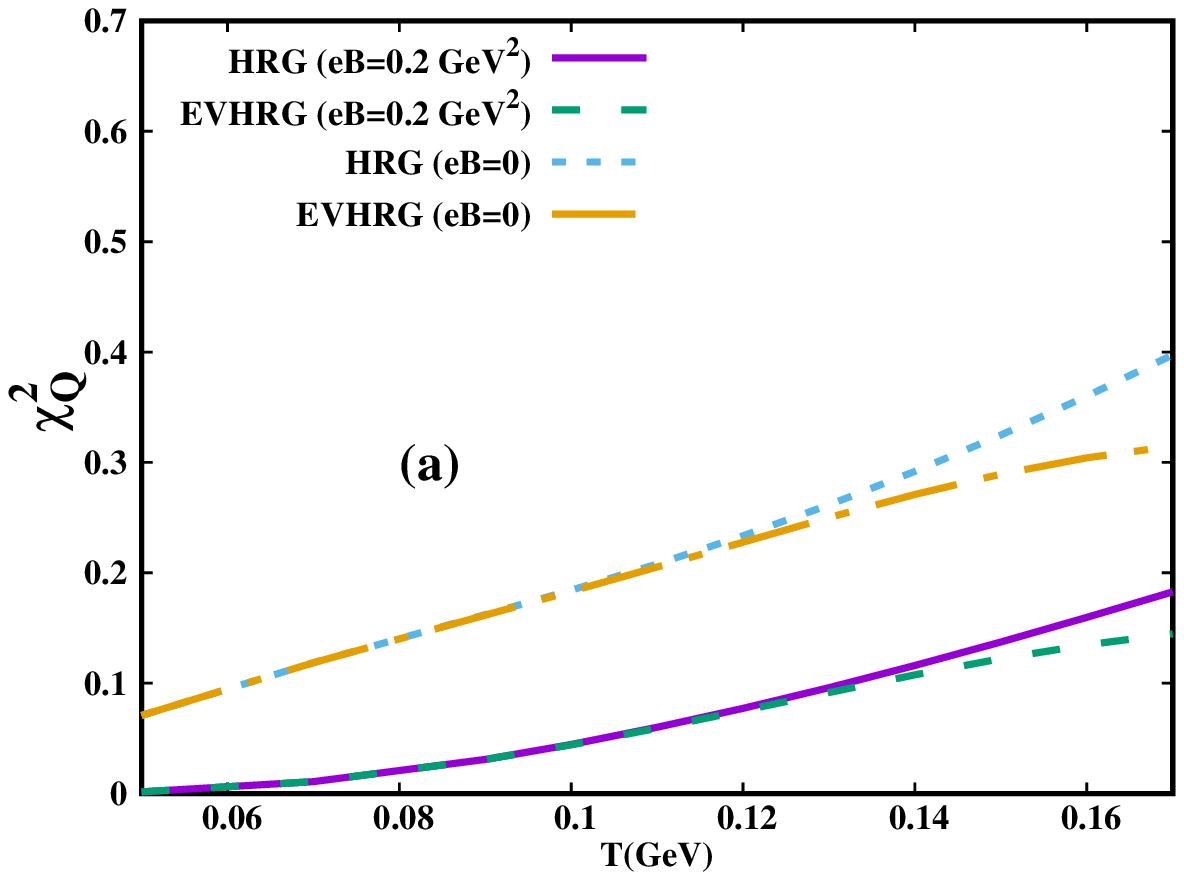}&
  \includegraphics[width=5cm,height=5cm]{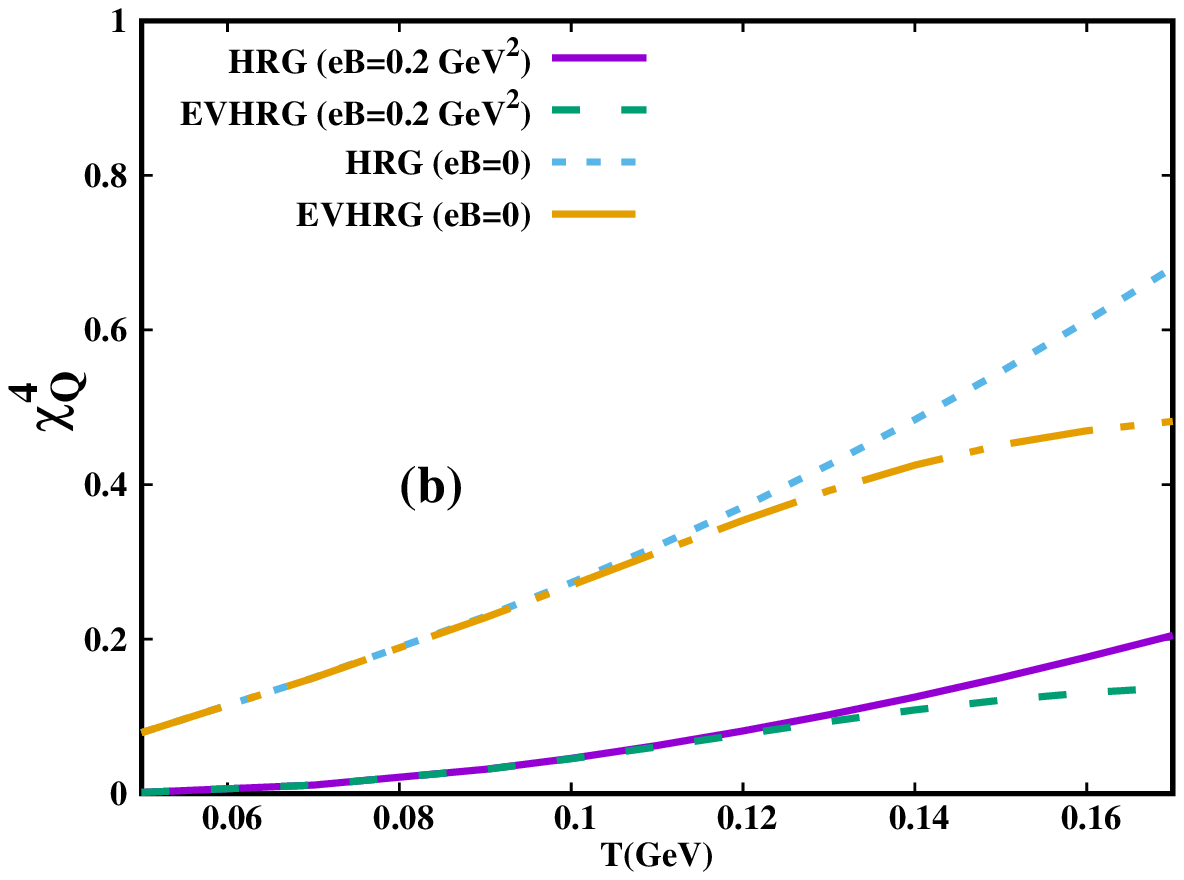}&
  \includegraphics[width=5cm,height=5cm]{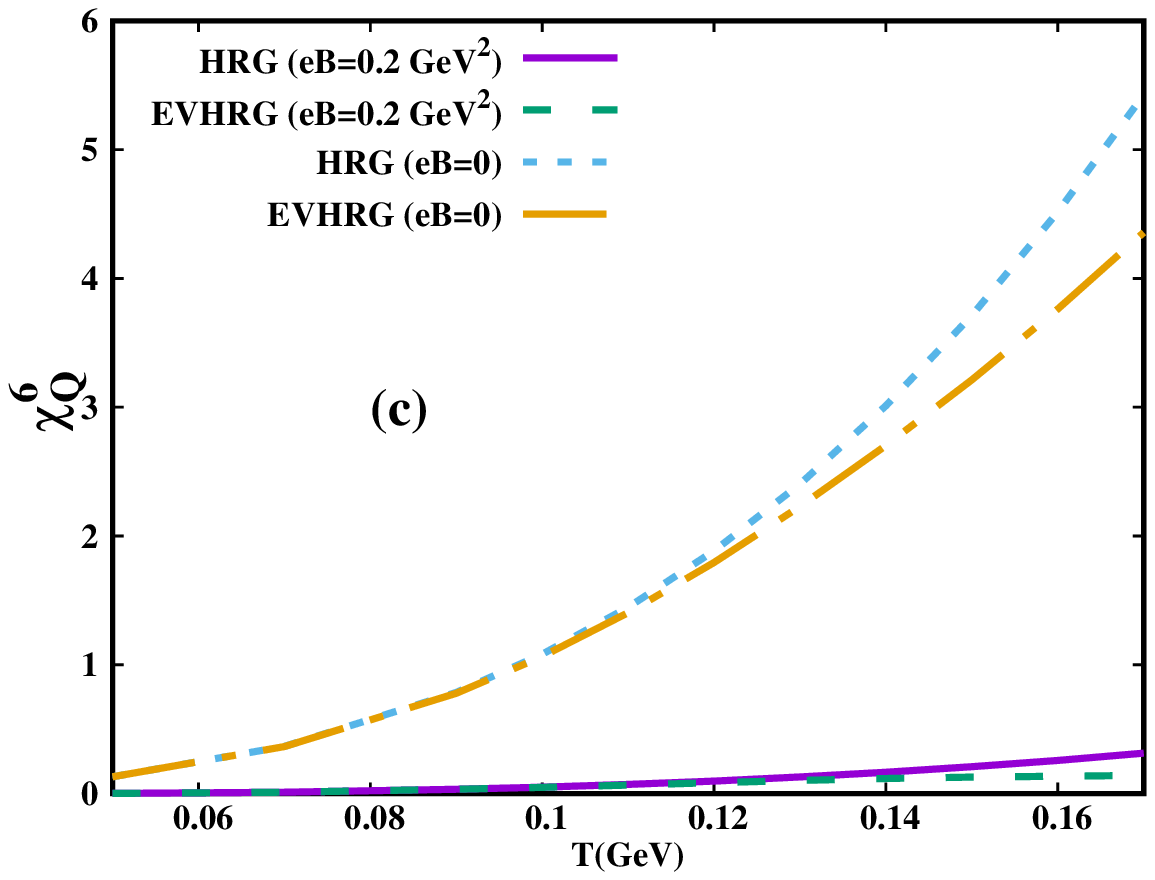}
  \end{tabular}
  \caption{(Color Online) Electric charge susceptibilities of the hadronic matter estimated using HRG and EVHRG with and without magnetic field. } 
\label{chiE}
  \end{center}
 \end{figure}

Fig.\ref{chiE} shows electric charge susceptibilities estimated within HRG and EVHRG models both in presence and in absence of magnetic field. In HRG model, electric charge susceptibilities of all orders are larger in magnitude than baryon susceptibilities. This is expected as in our particle spectrum there are  electrically charged mesons having very low mass, for example charged pions, which contribute to $\chi^{n}_{Q}$ unlike $\chi^{n}_{B}$. It is also important to note that higher order electric charge susceptibilities increase at a faster rate than lower ones. This is also due to lower masses of electrically charged hadrons. At low temperature, dominant contribution to $\chi_Q^n$s come from charged pions. As temperature increases, heavier hadrons start to contribute to pressure and hence to susceptibilities.  Here we have also taken into account the effect of Van der Waals excluded volume repulsive interaction.
The effect of hardcore repulsive interaction is to reduce pressure and hence $\chi_Q^n$ as compared to HRG Model. The effect of repulsive interaction is not significant at low temperatures ($T\sim 0.130 $GeV) since the population of hadrons with non-zero hard-core radius is low and we have chosen $r_\pi=0$ for pions which are dominant degrees of freedom. Thus the effect of repulsive interaction start to appear only after $T\sim 0.130 $ GeV when hadrons heavier than pions start to populate significantly.  We further note a  stronger suppression for higher order susceptibilities. It is seen that unlike baryon susceptibilities, there is a notable difference in magnitude between electric charge susceptibilities for $eB=0$  case and $eB=0.2$ GeV$^2$ case even at low temperatures. This observation can again be attributed to the fact that the lower mass electrically charged particles  are suppressed in presence of magnetic field as compared to $eB=0$ case.

For pure HRG one can calculate these susceptibilities analytically and can have have an understanding about this behaviour. The second and fourth order susceptibilities 
for pure HRG, with unit charge, can be written as

	\begin{equation*}
	\chi_i^2=\pm \sum_j\frac{g_j}{2\pi^2T^3}\int_{0}^{\infty}p^2 dp\frac{\exp{[(E_j-\mu_i)/T]}}{\{ \exp{[(E_j-\mu_i)/T]} \pm 1\}^2}
	\end{equation*}
	
	\begin{eqnarray}
		\chi_i^4&=\pm\sum_j \frac{g_j}{2\pi^2T^3}\int_{0}^{\infty}p^2 dp\left[\frac{\exp{[(E_j-\mu_i)/T]}}{\{ \exp{[(E_j-\mu_i)/T]} \pm 1\}^2} \right . \nonumber \\
		 &\left .- 6  \frac {(\exp{[E_j-\mu_i)/T])})^2} {\{ \exp{[(E_j-\mu_i)/T]} \pm 1\}^3} 
		+  6  \frac{(\exp{[E_j-\mu_i)/T])})^3}{\{ \exp{[(E_j-\mu_i)/T]} \pm 1\}^4} 
		 \right]
	\end{eqnarray}
	where $E_j$ is the energy of a particle of species $j$ and $\mu_i$ is chemical potential corresponding to some conserved charge.

From the above expressions one can find that $\chi^4_i$ has some extra terms compared to $\chi^2_i$. These terms are negligible at low temperatures if the mass of the 
particle is quite heavy compared to the temperature.   For the case of baryon,  $\chi_B^4$ and $\chi_B^2$  have almost same values  as the baryon charge 
is dominantly carried by protons which has a mass much higher than the temperature.  On the other hand, for electric charge, the leading order contribution comes from pion, the 
mass of which is comparable to the temperature. So all the terms in the expression of $\chi^4_Q$ contribute significantly and as a result  electric charge 
susceptibilities of different order have significantly different values.

  \begin{figure}[h]
	\vspace{-0.4cm}
	\begin{center}
		\begin{tabular}{c c}
			\includegraphics[width=7cm,height=7cm]{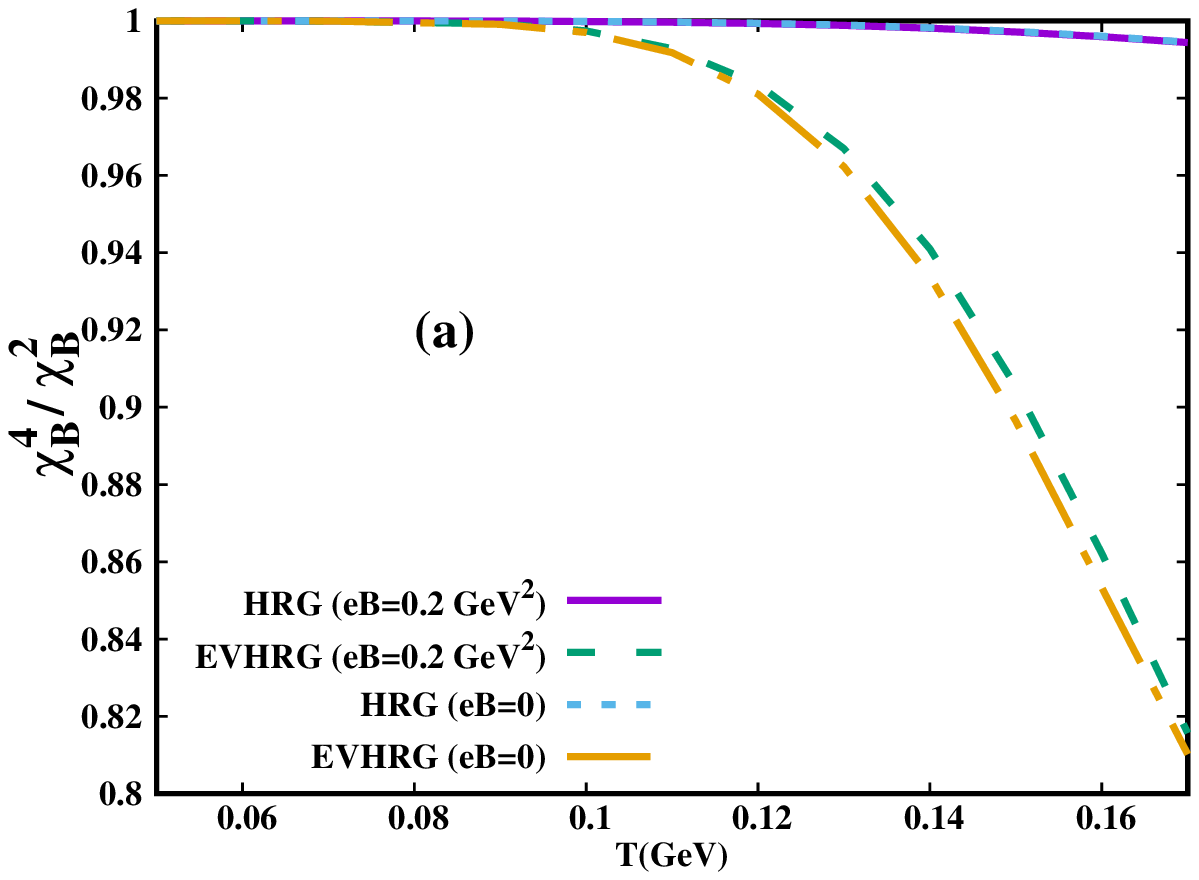}&
			\includegraphics[width=7cm,height=7cm]{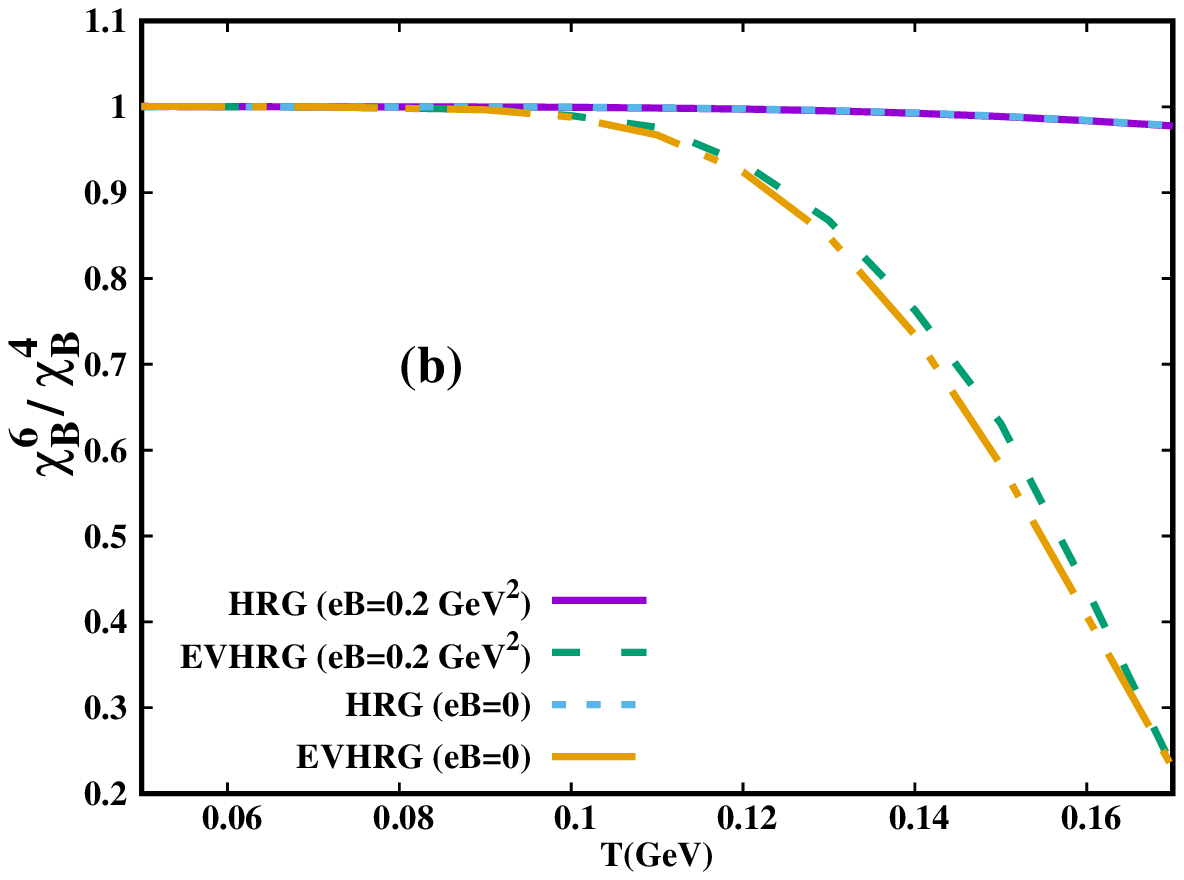}
		\end{tabular}
		\caption{(Color Online) Ratios of baryon susceptibilities of the hadronic matter estimated using HRG and EVHRG models with and without magnetic field. } 
		\label{chiB_ratios}
	\end{center}
\end{figure}

 \begin{figure}[b]
	\vspace{-0.4cm}
	\begin{center}
		\begin{tabular}{c c}
			\includegraphics[width=7cm,height=7cm]{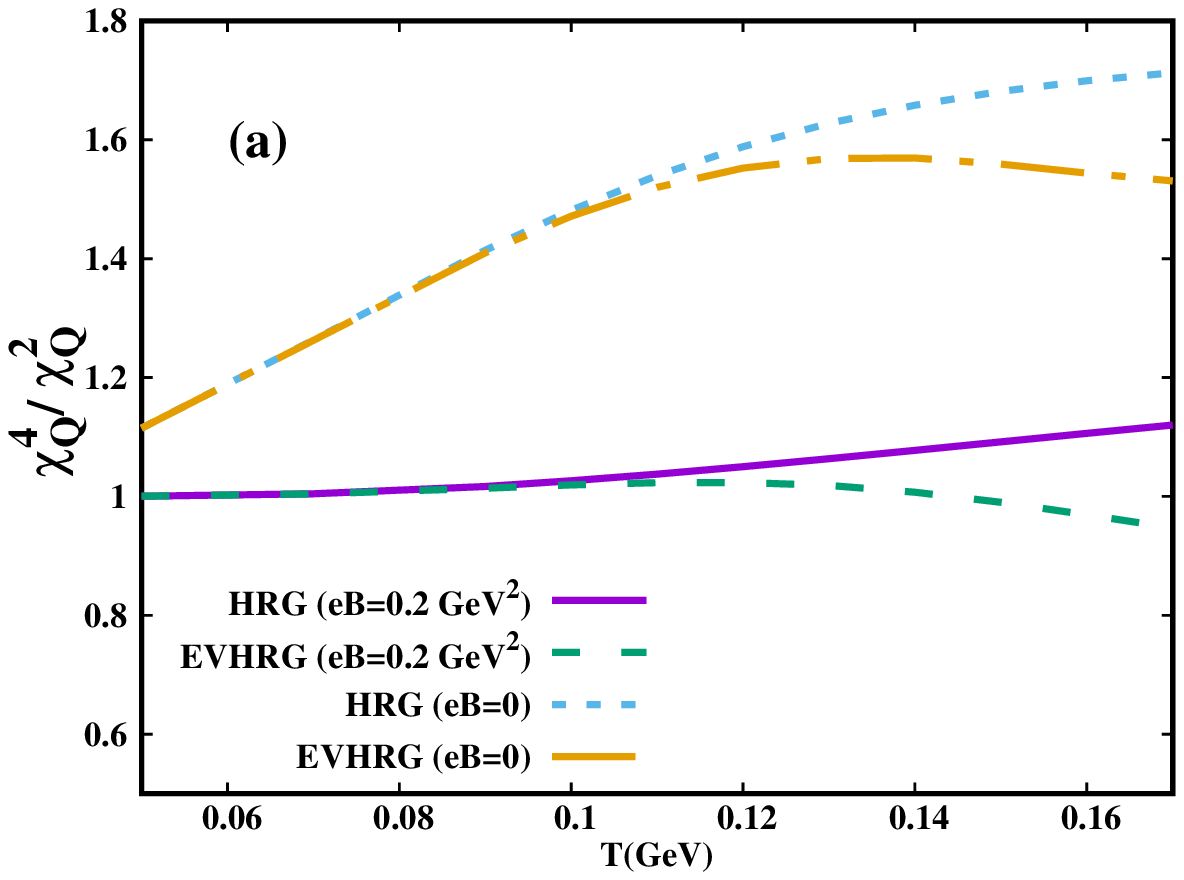}&
			\includegraphics[width=7cm,height=7cm]{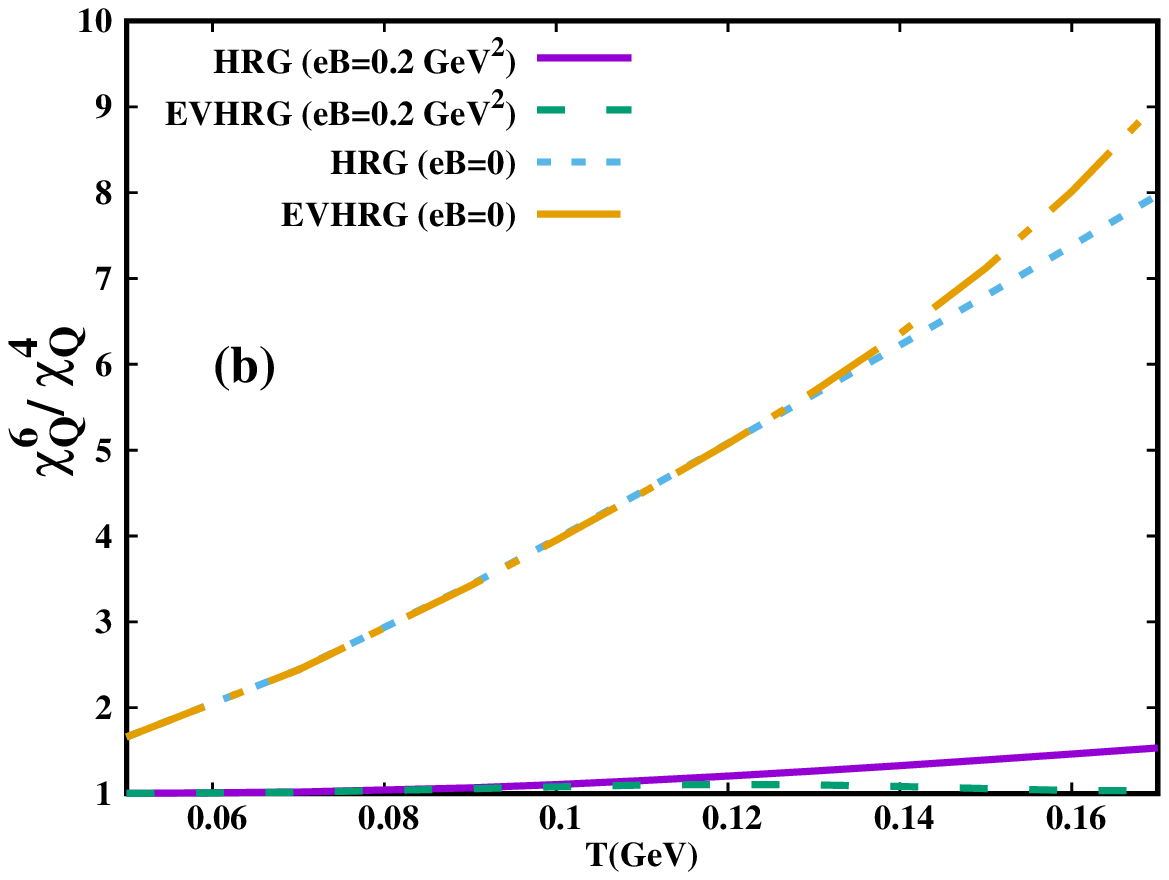}
		\end{tabular}
		\caption{(Color Online) Ratios of electric charge susceptibilities of the hadronic matter estimated using HRG and EVHRG models with and without magnetic field.} 
		\label{chiQ_ratios}
	\end{center}
\end{figure}

In figures \ref{chiB_ratios} and \ref{chiQ_ratios} we have plotted the ratios of susceptibilities. The ratio $\chi_B^4/\chi_B^2$ is almost unity throughout the temperature 
range for pure HRG model. For interacting scenario it decreases sharply by about $20\%$ at a temperature of $170 MeV.$  The ratio 
$\chi_B^6/\chi_B^4$, for pure HRG model, also stays close to unity throughout the temperature range.   
For interacting  scenario it decreases even faster than $\chi_B^4/\chi_B^2$. At a temperature of $170 MeV$ its value becomes almost $0.2$. 
The ratios of electric charge susceptibilities has a more complicated behaviour with temperature.  The ratio $\chi_Q^4/\chi_Q^2$ is close to unity at low 
temperature. For pure HRG without magnetic field it monotonically increases with temperature. For EVHRG without magnetic field it first increases with temperature 
and then starts decreasing. For the cases with magnetic field the changes are relatively small, at high temperature the change is about $10\%$. The ratio $\chi_Q^6/\chi_Q^4$ 
increases almost linearly with temperature for both HRG and EVHRG model with zero magnetic field. 
For finite magnetic field the ratio remains almost unity throughout the temperature range.

We now discuss the implications of our results in the context of heavy-ion collision experiments. 
Experimentally measured moments such as mean, standard deviation, skewness and kurtosis of conserved charges
are used to characterise the shape of charge distribution.  The products
of these moments are linked with susceptibilities. 
To make contact between our model and experimental data we need
a parametrisation of $T$, $\mu$'s and $B$ with $\sqrt{s}$ (centre of mass energy). Such
parametrisation exist for multiplicities of identified hadrons in the
absence of magnetic field~\cite{Cleymans_PRC_73,PLB695_Karsch} for
central collisions.  However no such parameterisation exists, till date, in presence 
of magnetic field. Secondly the parameters for central collisions
are not same as peripheral collisions. One may, at this stage, make a qualitative 
comparison~\cite{ab2}. However, a quantitative comparison is not possible till date.

Let us now discuss our results in the context of the available lattice findings. In Ref.~\cite{bali1} the  thermodynamic quantities have been calculated on the lattice. The authors of Ref.~\cite{bali1} have plotted pressure as a function of temperature for different values of the magnetic field. At $eB=0$, pressure estimated within the ambit  of both HRG and EVHRG model, as calculated in this work, are in both qualitative and quantitative agreement up to $T\sim 140$ MeV with the lattice data presented in Ref.~\cite{bali1}. At higher temperatures the qualitative match is very good whereas the exact values are little bit off.  At finite magnetic field also, both HRG and EVHRG results are in very good qualitative and quantitative agreement up to $T\sim 140$ MeV.  Once again, at higher temperatures the pressure calculated in the lattice framework is somewhat higher than that obtained in our model calculations. 
It would be also very  interesting to confront our results of baryon and electric charge susceptibilities with lattice QCD results once the lattice data for those observables are available. 
  \section{summary}
 \label{secVI}
  
 To summarise, we have analysed the effect of magnetic field on the hadronic matter using HRG and EVHRG models. The presence of magnetic field and the repulsive interactions, accounted through excluded volume corrections, significantly affect the static bulk thermodynamic quantities of hadronic matter. We observe that all the thermodynamical quantities are strongly suppressed due to non-zero background magnetic field and repulsive interactions. At finite $\mu_B$, the suppression, due to excluded volume, is even more than 
 that at zero $\mu_B$. The magnetic field affects the effective masses of hadrons thereby affecting the thermal population probability. While  effective mass of spin-0 particles increases in magnetic field, masses of spin-1 particles decreases. Upshot of this is the suppression of thermodynamic quantities. The repulsive interactions further enhance the suppression because of the finite size of hadrons which limits the available volume. The suppression of entropy density in presence of magnetic field and repulsive interaction is very important in the context of HIC experiments. We finally discussed the effect of magnetic field and repulsive interactions on the baryon number and electric charge susceptibility. We found that both of them are strongly suppressed at high temperature. Susceptibilities of all orders are 
 suppressed due to both magnetic field and repulsive interactions. The ratios of susceptibilities of different orders show dependence on excluded volume and magnetic field.

Thus, the effect of magnetic field cannot be ignored if one has to understand the low temperature hadronic matter. The importance of short range repulsive interactions in the context of heavy-ion collision experiment has already been established through  studies of dynamic as well as static properties of hadronic matter. In this work we tried to understand the effect of magnetic field along with the repulsive interactions on the static thermodynamic properties. If the magnetic field indeed exists in the hadronic phase of the matter produced in HICs then our study indicates that the magnetic field non-trivially affect the thermodynamics of hadronic matter. Effect of magnetic field will also be very important in the search of QCD critical point as we have seen in our study that the fluctuations of conserved charges are very non-trivially affected by magnetic field.

\section{Acknowledgement}

Guruprasad Kadam is financially supported by DST-INSPIRE Faculty research grant number DST/INSPIRE/04/2017/002293. A.B. thanks BRNS (DAE) and UGC (DRS) for 
support. AB also thanks Alexander von Humboldt (AvH) foundation and Federal Ministry of Education and Research (Germany) for support through Research Group Linkage programme. S.P. thanks  UGC for support.

 \appendix
 
\section{Formulae}
\label{formulae}

\begin{itemize}

 \item d-dimensional integral

 \begin{equation}
 \int_{-\infty}^{\infty}\frac{d^dp}{(2\pi)^d}\:(p^2+m^2)^{-A}=\frac{\Gamma(A-\frac{d}{2})}{(4\pi)^{d/2}\Gamma(A)(M^2)^{(A-\frac{d}{2})}}
 \label{dDimint}
 \end{equation}
 
 \item Riemann-Hurwitz $\zeta-$function
 
 \begin{equation}
\zeta(z,x)=\sum_{n=0}^{\infty}\frac{1}{(x+n)^z}
\label{RHdef}
\end{equation}

 with the expansion

 \begin{equation}
 \zeta\bigg(-1+\frac{\epsilon}{2},x\bigg)\approx-\frac{1}{12}-\frac{x^2}{2}+\frac{x}{2}+\frac{\epsilon}{2}\zeta^{'}(-1,x)+\mathcal{O}(\epsilon^2)
 \label{zetaexp}
 \end{equation}
 
 and the asymptotic behavior of the derivative

 \item The expansion of $\Gamma$-function is
 \begin{equation}
\Gamma\bigg(-1+\frac{\epsilon}{2}\bigg)=-\frac{2}{\epsilon}+\gamma-1+\mathcal{O}(\epsilon)
\label{gammaexp1}
\end{equation}

and 

 \begin{equation}
 \Gamma\bigg(-2+\frac{\epsilon}{2}\bigg)=\frac{1}{\epsilon}-\frac{\gamma}{2}+\frac{3}{4}+\mathcal{O}(\epsilon)
 \label{gammaexp2}
 \end{equation}
where $\gamma$ is the Euler constant.

\item The limiting expression for natural logarithm
 
 \begin{equation}
\lim_{\epsilon\longrightarrow 0}a^{-\epsilon/2}\approx 1-\frac{\epsilon}{2}\text{ln}(a)
\label{limit}
\end{equation} 
 
 \end{itemize}

\section{Renormalized $B$ dependent pressure for spin zero and spin one particles}
\label{vacpressure}

\begin{itemize}

 \item Spin-0 particle:
 \begin{equation}
\Delta P_{\text{vac}}^r(s=0,B)=-\frac{(eB)^2}{4\pi^2}\bigg(\zeta^{'}(-1,x+1/2)
-\frac{x^2}{2}\text{ln}(x)+\frac{x^2}{4}+\frac{\text{ln}(x)+1}{24}\bigg)
\label{delpvacBfin0}
 \end{equation}
 
  \item Spin-1 particle:
 \begin{eqnarray}
&&\Delta P_{\text{vac}}^r(s=1,B)=-\frac{3}{4\pi^2}(eB)^2\bigg(\zeta^{'}(-1,x-1/2)+\frac{(x+1/2)}{3}\text{ln}(x+1/2)\nonumber\\
&+&\frac{2}{3}(x-1/2)\text{ln}(x-1/2)
-\frac{x^2}{2}\text{ln}(x)+\frac{x^2}{4}-\frac{7}{24}(\text{ln}(x)+1)\bigg)
\label{delpvacBfin1}
 \end{eqnarray}
 
 \end{itemize}

\end{document}